# Versatile Photonic Encoding Platform with Selectable Back Ends for Automotive Edge Processing

**ISAMU TAKAI, MAKOTO NAKAI, AYANA MIZUNO, and ATSUTAKA MIYAMICHI**

Toyota Central Research and Development Laboratories, Incorporated, 41-1, Yokomichi, Nagakute, Aichi 480-1192, Japan

Corresponding author: Isamu Takai (takai@mosk.tytlabs.co.jp).

**ABSTRACT** The increasing diversity and computational demands of automotive edge functions have created a need for hardware architectures that support heterogeneous processing while limiting implementation complexity and system overhead. This work investigates a system-level approach in which a common, fixed photonic front end provides a shared physical transformation for multiple processing functions, while application adaptation is handled by lightweight peripheral processing. We propose the Versatile Photonic Encoding Platform (VPEP), implemented using a fabricated silicon-photonic chip with 32 phase-modulated inputs, a passive fixed encoding core, and 17 on-chip photodetector outputs. Without reconfiguring the photonic core, the same chip is experimentally evaluated across four representative functions relevant to automotive edge systems: classification and regression for machine learning, analog-to-digital conversion for mixed-signal processing, and photonic information encoding and recovery for communication-related processing. Compared with linear baselines without photonic encoding, VPEP improves the mean macro-F1 score on 12 of 13 classification datasets and the mean coefficient of determination on 9 of 12 regression benchmarks. For analog-to-digital conversion, the best configuration achieves an in-sample normalized root-mean-square error of approximately 0.75% of full scale. For information encoding and recovery, a lightweight nonlinear decoder reconstructs four-character (32-bit) inputs from 17-channel encoded outputs with approximately 94% character accuracy for a 26-symbol set. These results demonstrate the feasibility of cross-application processing with a single fixed silicon-photonic encoding core and suggest a system-level design approach, complementary to reconfigurable photonic architectures, that combines common photonic hardware with lightweight application-specific processing for future automotive edge systems.

**INDEX TERMS** Automotive electronics, edge computing, machine learning, optical computing, photonic integrated circuits, silicon photonics.

## I. INTRODUCTION

Advanced driver-assistance systems (ADAS) and automated-driving functions are becoming increasingly sophisticated, with more sensors and higher sensor data rates [1], [2]. As computational and communication demands on in-vehicle systems continue to grow, automotive electrical/electronic (E/E) architectures are evolving toward centralized computing and zonal architectures, while sensors and edge-processing modules remain physically distributed throughout the vehicle [3]–[5]. Performing selected processing functions near the sensors can reduce network traffic and central computing load while supporting low-latency operation [6]–[8]. However, as the diversity of edge functions increases, implementing them with separate task-specific electronic circuits or processing blocks can lead to hardware fragmentation and increased design, integration, and validation effort [9], [10]. At the same time, the increasing computational complexity of these functions places greater demands on power and processing resources [11], [12]. A key system-level challenge is therefore to support continued growth in vehicle functionality while limiting increases in hardware complexity and development cost.

A common front-end architecture that performs a shared signal transformation for multiple downstream tasks could reduce this fragmentation. Because it lies on the signal path of every downstream function, such a front end should process many input channels in parallel with high throughput and low implementation overhead. Photonics is well suited to these requirements because optical propagation and interference can mix these channels simultaneously without sequential electronic arithmetic. Silicon photonics further enables these optical functions to be integrated into a compact chip-scale platform [13], [14]. Together, these properties make silicon photonics a promising physical platform for a common high-bandwidth front end, with the potential for lower energy per transformation and faster front-end processing [15], [16].

Motivated by these potential advantages, integrated photonic computing has attracted substantial interest, particularly for neural-network and matrix-processing applications [16]. One major approach employs programmable photonic architectures, in which optical weights or circuit functions are reconfigured for individual tasks [17]–[20]. Such programmability provides high task-specific flexibility; however, scaling the number of tunable parameters can increase device area, control complexity, calibration requirements, and, depending on the tuning mechanism, static power consumption [21]–[23]. These implementation considerations are particularly important for automotive deployment, where hardware must meet stringent requirements for cost efficiency, reliability, and long-term operation [11]. An alternative approach is to keep the physical photonic transformation fixed and train only a lightweight electronic readout, as in physical reservoir computing and extreme learning machines [24]–[26]. By reducing the need for task-programmable photonic elements and associated control circuitry, this fixed approach is well aligned with automotive requirements that emphasize hardware simplicity and robustness.

Although fixed photonic approaches have been investigated for a variety of computational tasks [24]–[26], experimental demonstrations spanning fundamentally different application domains remain scarce. Previous multi-task photonic reservoir studies, for instance, experimentally evaluated two prediction tasks [27], whereas a broader range of learning and signal-processing tasks was investigated only through numerical simulation [28]. More broadly, automotive applications have rarely been the primary focus of previous photonic computing studies. To the best of our knowledge, no previous study has systematically evaluated the same fabricated photonic device across fundamentally different application domains relevant to automotive edge processing without reconfiguring its photonic core. Such experimental validation is a necessary step toward establishing whether a fixed photonic core can serve as a practical common front end for heterogeneous automotive functions while limiting hardware complexity and system overhead.

To address this gap, we propose a Versatile Photonic Encoding Platform (VPEP), which is based on a common silicon-photonic front end and selectable application-specific electronic back ends. The VPEP chip comprises a 32-channel phase-modulator (PM) array at the input, a passive photonic encoding core, and a 17-channel on-chip photodetector (PD) array. The photonic encoding core implements a fixed optical mixing transformation and remains unchanged across applications. Electrical inputs modulate the phases of optical signals, which are then transformed through passive propagation and interference and converted into encoded electrical outputs by photodetection. Application-specific lightweight electronic back ends are then used to adapt the common photonic front end to different functions. Our previous work established the underlying optical encoding mechanism and its relationship to random Fourier feature mapping for nonlinear function approximation [29]. Whereas that work focused on the operating principle and nonlinear function approximation enabled by the photonic encoder, the present study investigates the cross-application versatility of the same fixed photonic hardware across fundamentally different processing functions relevant to automotive edge applications.

We experimentally evaluate the same VPEP hardware across four representative processing functions spanning three application domains: classification and regression for machine learning, analog-to-digital (A/D) conversion for mixed-signal processing [30], [31], and photonic information encoding and recovery for communication-related processing. For classification and regression, we evaluate whether the fixed photonic transformation can provide useful feature representations for lightweight linear back ends across diverse public datasets. The A/D conversion experiment combines VPEP encoding with per-channel quantization and lightweight digital decoding. The information-encoding experiment evaluates whether 32-bit digital inputs can be transformed by the same fixed VPEP front end into 17-channel encoded outputs and subsequently recovered using lightweight electronic decoders. This device-dependent transformation provides a potential basis for future applications involving dimensionality reduction, compact information representation, data obfuscation, and physical-layer security [32], [33]. The objective is not to optimize the photonic hardware separately for each function, but to evaluate whether the same fixed photonic encoding core can serve as a common front end across these substantially different processing functions.

The main contributions of this work are as follows:

1. **Versatile photonic architecture:** We introduce VPEP, which separates a common, fixed silicon-photonic encoding front end from application-specific lightweight electronic back ends.
2. **Integrated hardware implementation:** We realize VPEP using a silicon-photonic chip that integrates a 32-channel PM array, a compact passive interference-based encoding core, and a 17-channel on-chip PD array.
3. **Cross-application experimental evaluation:** Using the same fabricated photonic hardware without task-specific reconfiguration of the photonic encoding core, we experimentally demonstrate classification, regression, A/D conversion, and information encoding and recovery, and evaluate the applicability and limitations of fixed photonic encoding across heterogeneous edge-processing functions.

These results demonstrate the feasibility of VPEP as a common photonic encoding framework across heterogeneous applications and provide a basis for investigating how fixed photonic transformations can be incorporated into future automotive and other distributed edge-processing systems.

## II. VERSATILE PHOTONIC ENCODING PLATFORM

### A. VPEP CONCEPT AND IN-VEHICLE DEPLOYMENT SCENARIO

Fig. 1 illustrates the concept of the VPEP and an envisioned in-vehicle deployment scenario. In future centralized and zonal automotive E/E architectures, sensors and edge-processing modules are expected to remain physically distributed throughout the vehicle [3]–[5]. Such edge modules may perform diverse functions, including signal conditioning, A/D conversion, local inference, and communication preprocessing. VPEP is intended to provide a common photonic front end for these heterogeneous processing functions.

In each edge module, the VPEP photonic encoding core performs a fixed optical mixing transformation through passive propagation and interference. Electrical inputs $\boldsymbol{u}$ modulate the phases of optical signals, which are then transformed by the fixed photonic core and converted into encoded electrical outputs $\boldsymbol{z}$ through photodetection. Importantly, the photonic core itself is not reconfigured for individual applications; application-dependent adaptation is handled outside the fixed core using peripheral input mapping and lightweight electronic back-end processing. Because the optical transformation is realized through passive propagation and interference, VPEP can exploit the intrinsic parallelism and high bandwidth of photonic processing, offering the potential for lower energy per transformation and faster front-end processing, subject to system-level constraints and interface overheads [15], [16]. The detailed operating principle and prototype implementation are described in Section II-B.

This separation between a common fixed photonic core and application-dependent peripheral processing allows the same fabricated VPEP hardware to be applied across heterogeneous automotive edge functions. For example, encoded outputs generated by the fixed photonic core can be processed by lightweight electronic back ends for classification or regression, combined with quantization and digital decoding for A/D conversion, or used for information encoding and recovery with receiver-side decoding. By maintaining a common photonic front end across these functions, VPEP provides a system-level approach for reducing hardware fragmentation and repeated hardware design while supporting multiple edge-processing applications.

The in-vehicle deployment shown in Fig. 1 represents a longer-term application scenario. It illustrates the broader concept pursued in this work: a fixed photonic encoding core serving as a common hardware primitive that can be combined with application-specific peripheral processing to support evolving automotive edge functions.

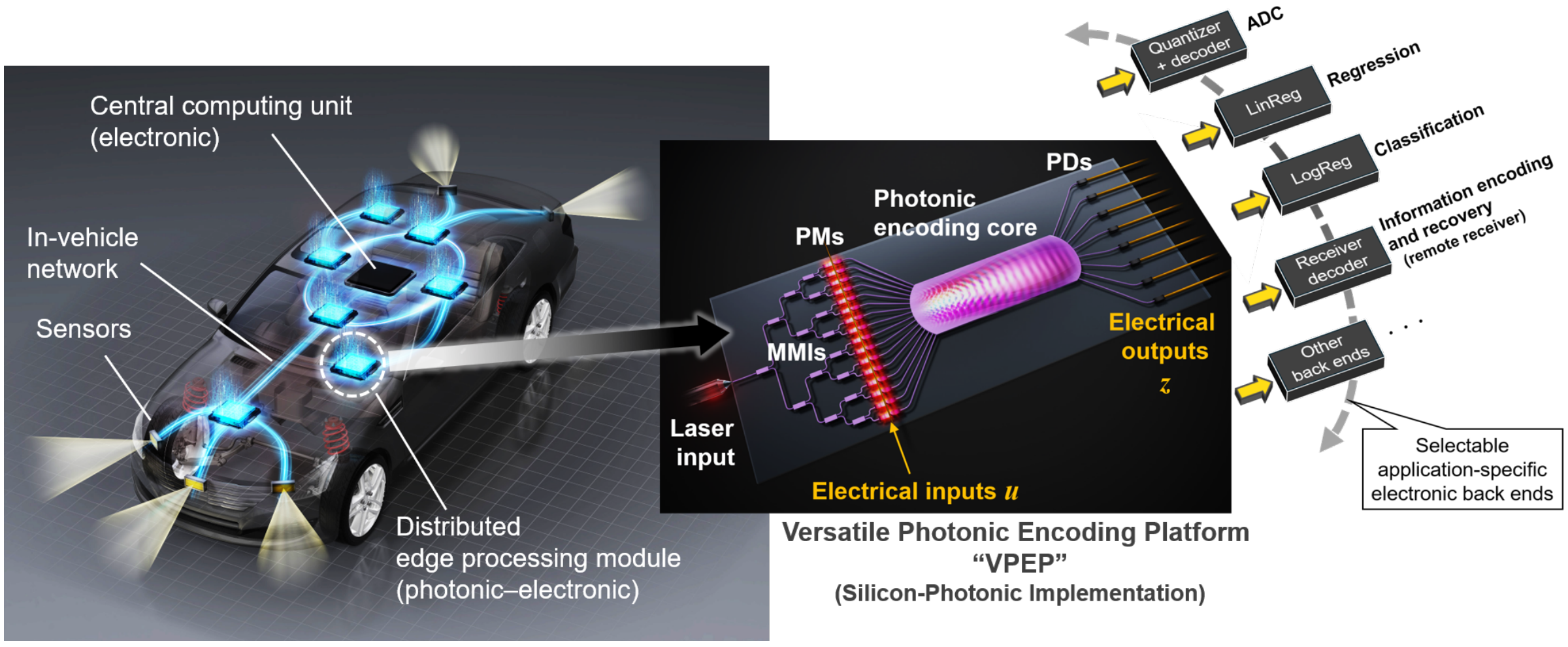

FIGURE 1. VPEP concept and envisioned in-vehicle deployment scenario. Distributed photonic–electronic edge-processing modules employ a common silicon-photonic front end with a fixed photonic encoding core that maps electrical inputs $\boldsymbol{u}$ to encoded electrical outputs $\boldsymbol{z}$. The outputs are processed by selectable application-specific electronic back ends to support heterogeneous edge-processing functions. PMs: phase modulators, PDs: photodetectors, MMIs: multimode interferometers, ADC: analog-to-digital converter.

### B. OPERATING PRINCIPLE AND PROTOTYPE CHIP

Fig. 2(a) summarizes the signal-flow model of VPEP. For clarity, the photonic encoding core refers specifically to the passive interference region that implements the fixed optical mixing matrix $H$. The VPEP chip comprises a 32-channel PM array for electrical-to-optical (E/O) conversion, the photonic encoding core, and a 17-channel on-chip PD array for optical-to-electrical (O/E) conversion. Together, these elements map the electrical input vector $\boldsymbol{u}$ to the encoded electrical output vector $\boldsymbol{z}$. The VPEP chip is preceded by an input mapper and followed by a selectable application-specific electronic back end.

The input $x$, which may be a scalar or vector, is first mapped to a 32-channel voltage vector $\boldsymbol{u} \in \mathbb{R}^{32}$ by the input mapper,

$$\boldsymbol{u} = M(x). \qquad (1)$$

The input mapper $M(\cdot)$ and electronic back end $B_{\text{task}}(\cdot)$ provide application-dependent preprocessing and postprocessing, respectively, while the photonic transformation implemented by the VPEP chip remains fixed.

Input laser light is divided into 32 optical channels by an on-chip 1×2 multimode-interference (MMI) splitter tree and routed to the PM array. The present prototype employs thermo-optic PMs. The voltage vector $\boldsymbol{u}$ drives the PM array and produces a complex optical field vector $\boldsymbol{a} \in \mathbb{C}^{32}$. This mapping is denoted by $\boldsymbol{a} = \varphi(\boldsymbol{u})$ in Fig. 2(a). The $n$-th modulated optical field is modeled as

$$a_n = A \exp\{ j(\alpha_n u_n^2 + \beta_n)\}, \qquad (2)$$

where $A$ is the optical field amplitude, $\alpha_n$ is the squared-voltage-to-phase coefficient, $\beta_n$ is a static phase offset, and $j$ denotes the imaginary unit. For simplicity, a common amplitude $A$ is assumed; channel-dependent optical loss can be incorporated by replacing $A$ with a channel-dependent amplitude $A_n$.

The photonic encoding core then performs a fixed optical mixing transformation through passive propagation and interference. This transformation is modeled by a complex matrix $H \in \mathbb{C}^{17\times 32}$,

$$\boldsymbol{r} = H\boldsymbol{a}, \qquad (3)$$

where $\boldsymbol{r} \in \mathbb{C}^{17}$ represents the optical fields at the 17 output channels. The matrix $H$ is determined by the fabricated photonic structure and remains unchanged across applications.

The output optical fields are detected by the on-chip PDs, producing the encoded electrical output vector $\boldsymbol{z} \in \mathbb{R}^{17}$, with $z_i \propto |r_i|^2$.

In our previous work, the combination of optical phase modulation, fixed optical mixing, and square-law photodetection was theoretically and experimentally investigated in relation to random Fourier feature mapping for nonlinear function approximation [29]. In this framework, optical interference followed by photodetection produces a nonlinear representation of the input that can be exploited by subsequent electronic processing. Detailed theoretical analysis and experimental demonstrations of nonlinear function approximation are provided in [29]. In the present VPEP framework, the resulting encoded output $\boldsymbol{z}$ is processed by an application-specific electronic back end,

$$y = B_{task}(\boldsymbol{z}). \qquad (4)$$

Thus, the same fixed photonic front end can generate encoded representations for use across different applications, while application-dependent adaptation is handled by the surrounding input mapper and electronic back end.

Fig. 2(b) shows optical micrographs of the fabricated silicon-photonic VPEP chip and a magnified view of its photonic encoding core. The core consists of a compact passive interference region connecting 32 input waveguides to 17 output waveguides and physically implements the device-defined optical mixing described above. In the present prototype, the interference region measures approximately 33 μm in width and 147 μm in length.

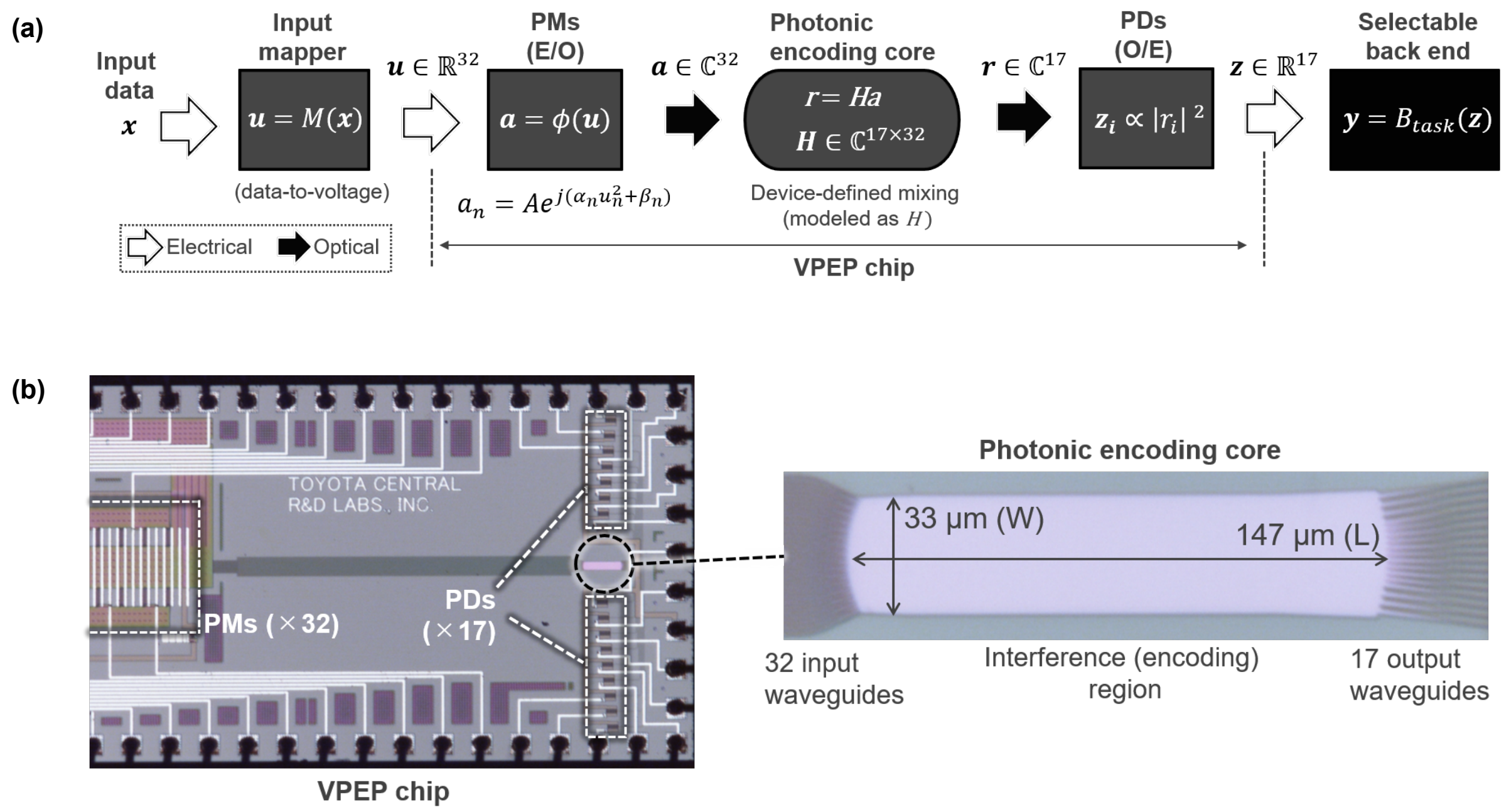

**FIGURE 2. Signal flow model and fabricated device of silicon-photonic VPEP. (a) Signal flow model comprising an input mapper, the VPEP chip, and a selectable application-specific electronic back end. The VPEP chip consists of a 32-channel phase-modulator (PM) array, a fixed photonic encoding core, and a 17-channel photodetector (PD) array. White and black arrows denote electrical and optical signals, respectively. (b) Optical micrographs of the fabricated VPEP chip and a magnified view of the photonic encoding core, showing the passive interference region with 32 input and 17 output waveguides. E/O: electrical/optical, O/E: optical/electrical.**

## III. APPLICATION DEMONSTRATIONS

### A. EXPERIMENTAL SETUP AND COMMON PROCESSING FLOW

Fig. 3(a) illustrates the experimental configuration used to evaluate VPEP across multiple applications. The same VPEP photonic front end is combined with application-specific input mapping and lightweight electronic back-end processing. A continuous-wave (CW) laser at 1550 nm provides the optical carrier, which is coupled into the VPEP chip through a polarization controller (Pol. Ctrl.) and an optical attenuator (ATT.). The VPEP chip is mounted on a printed circuit board (PCB) and integrates the PM array, photonic encoding core, and on-chip PD array, as described in Section II-B. During all experiments, the chip temperature is maintained at 35 °C using a Peltier temperature controller.

The input $x$ is mapped to a 32-dimensional drive vector $\boldsymbol{u} \in \mathbb{R}^{32}$ by the input mapper. As detailed in Fig. 3(b), the mapper performs dimensionality adaptation via replication and zero padding, followed by channel assignment and voltage scaling, thereby adapting application-dependent input formats to the fixed 32-channel VPEP interface. The drive vector $\boldsymbol{u}$ is generated digitally on the PC and converted into 32 analog drive voltages by a 32-channel, 16-bit digital-to-analog converter (DAC). These voltages drive the PM array, and the resulting phase-modulated optical signals propagate through the fixed photonic encoding core. After on-chip photodetection, the 17-channel analog encoded outputs $z \in \mathbb{R}^{17}$ are digitized by an 18-bit analog-to-digital converter (ADC). The DAC and ADC are integrated into the data-acquisition (DAQ) system, enabling synchronized PM driving and PD readout.

In the present experiments, both the input mapper and the application-specific electronic back end are implemented in software on a PC. The input mapper generates the drive vector applied through the DAC, whereas the back end processes the digitized encoded outputs $z$ to produce the application-dependent result, $y = B_{\text{task}}(z)$.

Fig. 3(b) details the input mapper $\boldsymbol{u} = M(x)$, which provides a lightweight interface between application-dependent input formats and the fixed 32-channel input dimensionality of VPEP. Dimensionality adaptation is first performed via replication and zero padding to obtain $\tilde{x} \in \mathbb{R}^{32}$,

$$\tilde{x} = \mathrm{pad}_{32}(\mathrm{rep}_\rho(x)), \tag{5}$$

where $\mathrm{rep}_\rho(\cdot)$ denotes replication with replication factor $\rho$ and $\mathrm{pad}_{32}(\cdot)$ denotes zero padding to 32 dimensions. Channel assignment is then performed using a permutation matrix $P \in \{0,1\}^{32\times32}$, followed by voltage scaling $f_V(\cdot)$, yielding

$$\boldsymbol{u} = f_V(P\tilde{x}). \tag{6}$$

The voltage-scaling function $f_V(\cdot)$ maps the channel-assigned input values to PM drive voltages to produce the corresponding optical phase shifts. The replication factor $\rho$ and permutation matrix $P$ determine how the input information is distributed across the 32 modulation channels. Because the VPEP response depends on the relative optical phases generated across these channels, changing the mapper configuration alters the interference condition in the photonic encoding core and consequently the encoded output $z$. The resulting application performance can therefore depend on the input mapper configuration. A general task-agnostic criterion for selecting $\rho$ and $P$ has not yet been established. Accordingly, a finite set of candidate configurations is examined in a preliminary evaluation, after which the selected configuration is held fixed during the reported evaluation. Developing a task-agnostic input mapping strategy remains an important direction for future work.

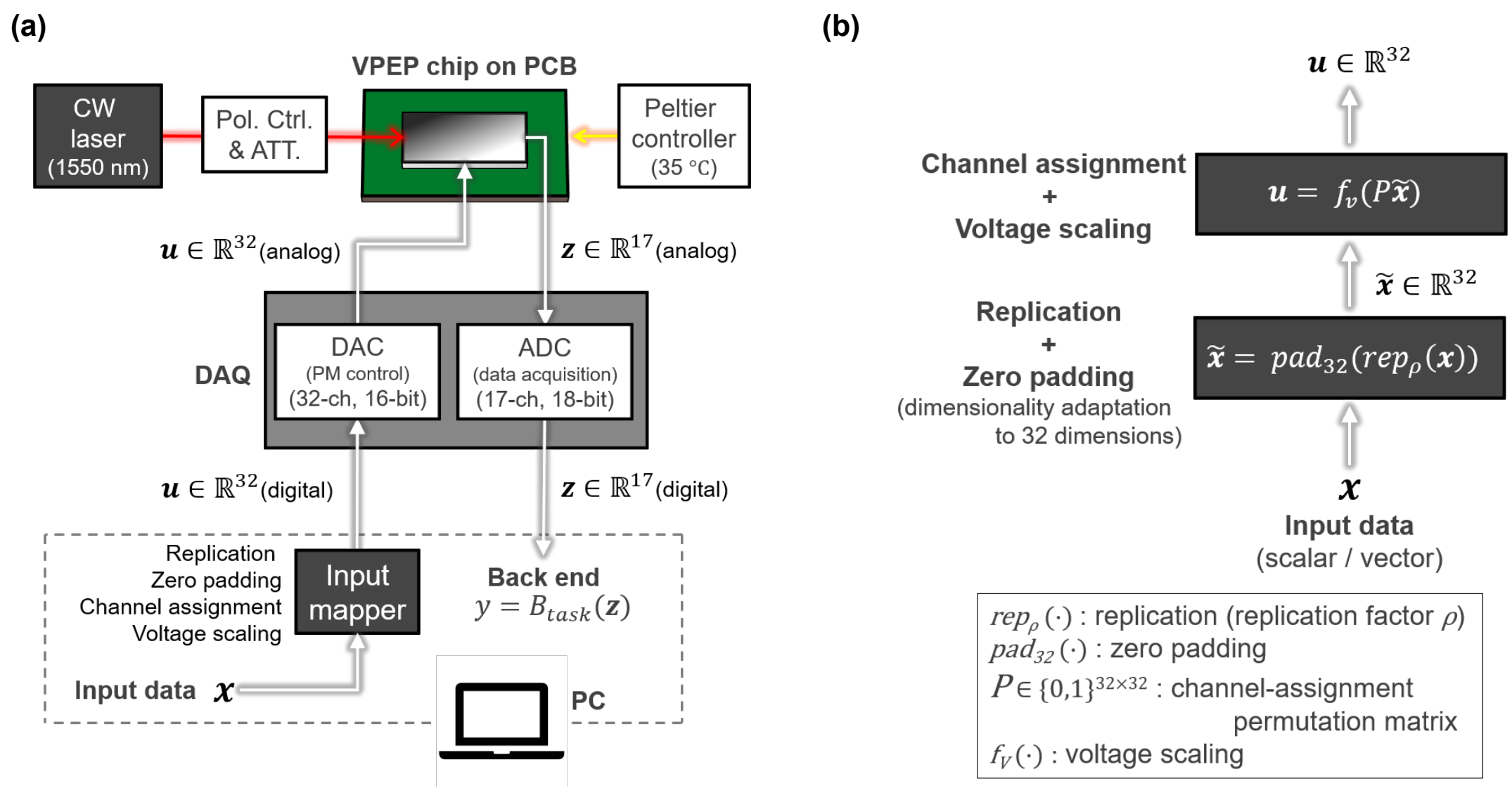

**FIGURE 3. Experimental setup and input mapping flow for VPEP evaluation. (a) Experimental setup comprising a 1550-nm continuous-wave (CW) laser, polarization controller (Pol. Ctrl.), optical attenuator (ATT.), VPEP chip on a PCB maintained at 35 °C by a Peltier controller, a 32-channel 16-bit digital-to-analog converter (DAC), a 17-channel 18-bit analog-to-digital converter (ADC), and PC-based input mapping and back-end processing. (b) Input mapper $\boldsymbol{u} = M(\boldsymbol{x})$, comprising dimensionality adaptation via replication and zero padding, channel assignment, and voltage scaling.**

### *B. CLASSIFICATION AND REGRESSION*

#### 1) EVALUATION PROTOCOL

Fig. 4 summarizes the evaluation pipelines for the classification and regression tasks. Unless otherwise noted, the VPEP hardware and measurement configuration are identical to those described in Section III-A. The objective of this evaluation is to characterize the feature-transformation capability of the fixed VPEP front end across diverse datasets when combined with lightweight electronic back ends, rather than to maximize task performance using computationally intensive models.

Figs. 4(a) and 4(b) show the classification and regression pipelines, respectively. In both cases, the input feature vector $\boldsymbol{x}$ is converted by the input mapper into the 32-channel drive vector $\boldsymbol{u}$ and applied to the VPEP front end. The measured encoded output vector $z$ is then standardized and processed by a lightweight electronic back end: logistic regression (LogReg) for classification and linear regression (LinReg) for regression. Both back ends are implemented using scikit-learn [34] with the LogisticRegression [35] and LinearRegression [36] classes, respectively. The back-end models are trained in a supervised manner using the ground-truth targets $y_t$. LogReg is trained with L2 regularization, whereas LinReg is fitted by ordinary least squares.

The experimental conditions are summarized in Table I. Feature-wise min-max scaling to [0, 1] is applied before input mapping to match the operating range of the VPEP interface. Classification performance is evaluated using repeated stratified $k$-fold cross-validation (CV) with $k = 4$ or 5 depending on the dataset and 10 repeats, implemented with RepeatedStratifiedKFold [37]. Regression performance is evaluated using repeated five-fold CV with 10 repeats, implemented with RepeatedKFold [38]. Classification is assessed primarily using the macro-F1 score, defined as the unweighted mean of the per-class F1 scores and computed using f1_score [39]; accuracy is reported as a supplementary metric. Regression is assessed primarily using the coefficient of determination ($R^2$), computed using r2_score [40]; root-mean-square error (RMSE) is reported as a supplementary metric. The replication factor $\rho$ and channel assignment $P$ are selected for each dataset through preliminary evaluation and are subsequently held fixed throughout the reported cross-validation evaluation.

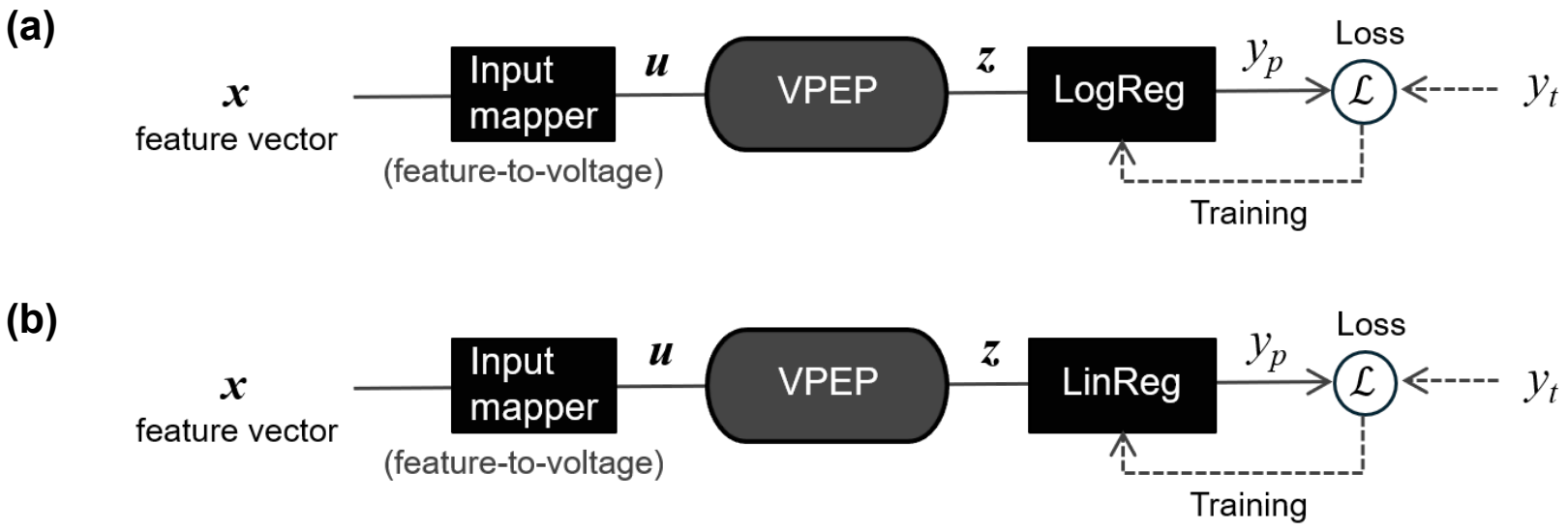

**FIGURE 4. Evaluation pipelines for classification and regression. (a) Classification pipeline with logistic regression (LogReg). (b) Regression pipeline with linear regression (LinReg).**

TABLE I
EXPERIMENTAL CONDITIONS FOR CLASSIFICATION AND REGRESSION

| | Classification | Regression |
|---|---|---|
| **Algorithm** | LogReg | LinReg |
| **Training objective** | Log loss | Squared error |
| **Input data $x$** | Feature vector: from the UCI Machine Learning Repository or scikit-learn built-in datasets | |
| **Data normalization** | Feature-wise min–max scaling to [0,1] | |
| **Evaluation protocol** | Repeated stratified $k$-fold CV | Repeated $k$-fold CV |
| **Folds / repeats** | $k$ = 4 or 5, repeats = 10 | $k$ = 5, repeats = 10 |
| **Replication factor $\rho$** | Dataset-dependent; see Tables S-I and S-II | |
| **Channel assignment (permutation $P$)** | Dataset-dependent; determined by preliminary evaluation | |
| **Performance metrics** | macro-F1, accuracy (see Table S-I) | $R^2$, RMSE (see Table S-II) |

LogReg: logistic regression; LinReg: linear regression; CV: cross-validation; $R^2$: coefficient of determination; RMSE: root-mean-square error. For the Zoo dataset, $k = 4$ because the number of samples in a minority class is insufficient for five-fold stratified CV.

### 2) RESULTS: CLASSIFICATION

Classification performance is evaluated by comparing the proposed VPEP + LogReg pipeline with the LogReg baseline using standardized input features without photonic encoding across 13 public datasets spanning 3–30 input features and 2–10 classes. The datasets are drawn from the UCI Machine Learning Repository [41] and the scikit-learn built-in collections [34]. This diverse benchmark set enables evaluation of whether the same fixed VPEP hardware can provide useful feature transformations across classification problems with different input dimensionalities, feature distributions, and class structures. The following datasets are used: Land Mines-1 (LM) [42], [43], Iris [44], Thyroid (New) (TR) [45], Wholesale Customers (WC) [46], Wireless Indoor Localization (WIL) [47], Yeast [48], Glass Identification (GI) [49], MAGIC Gamma Telescope (MG) [50], Heart Failure Clinical Records (HFCR) [51], Wine [52], Heart Disease (processed Cleveland) (HD) [53], Zoo [54], and Breast Cancer Wisconsin (Diagnostic) (BC) [55].

Fig. 5(a) compares the mean macro-F1 scores of the baseline and VPEP + LogReg pipelines in a parity plot. VPEP + LogReg achieves a higher mean macro-F1 score than the baseline on 12 of the 13 datasets, demonstrating that the fixed photonic transformation improves classification performance across a broad range of the evaluated datasets when paired with the same lightweight linear classifier. Particularly large improvements are observed for MG, GI, LM, and WC, with macro-F1 gains of approximately 0.12–0.23. For datasets such as Iris, Wine, and WIL, where the baseline already achieves high macro-F1 scores, the changes are smaller; VPEP + LogReg nevertheless achieves a macro-F1 score of 1.0 for Iris. BC is the only dataset for which the mean macro-F1 score decreases, by 0.0035, a difference that is small relative to the observed cross-validation variability.

Although the magnitude of the improvement varies among datasets, no simple dataset-level property has yet been identified that reliably predicts the benefit of VPEP encoding. Full numerical results, including accuracy, macro-F1, and cross-validation statistics, are provided in Table S-I.

To examine how VPEP encoding changes the feature representation, Figs. 5(b)–(e) show representative linear discriminant analysis (LDA) projections [56] for Iris and Thyroid (New), computed using scikit-learn's LinearDiscriminantAnalysis [57]. LDA is used solely for visualization and is not part of the classification pipeline. The baseline projections are obtained from the original input features $\boldsymbol{x}$, whereas the VPEP projections are obtained from the encoded features $\boldsymbol{z}$. In both examples, the VPEP-encoded features exhibit more clearly separated class clusters than those for the corresponding baseline representations, indicating improved linear class separability after photonic encoding. This observation is consistent with the increased Fisher eigenvalue sums shown in Figs. 5(b)–(e), which provide a quantitative measure of class separation based on the Fisher discriminant framework [56]. These results support the interpretation that the device-defined photonic transformation can reshape the input feature space into a representation that is more amenable to classification by a lightweight linear back end.

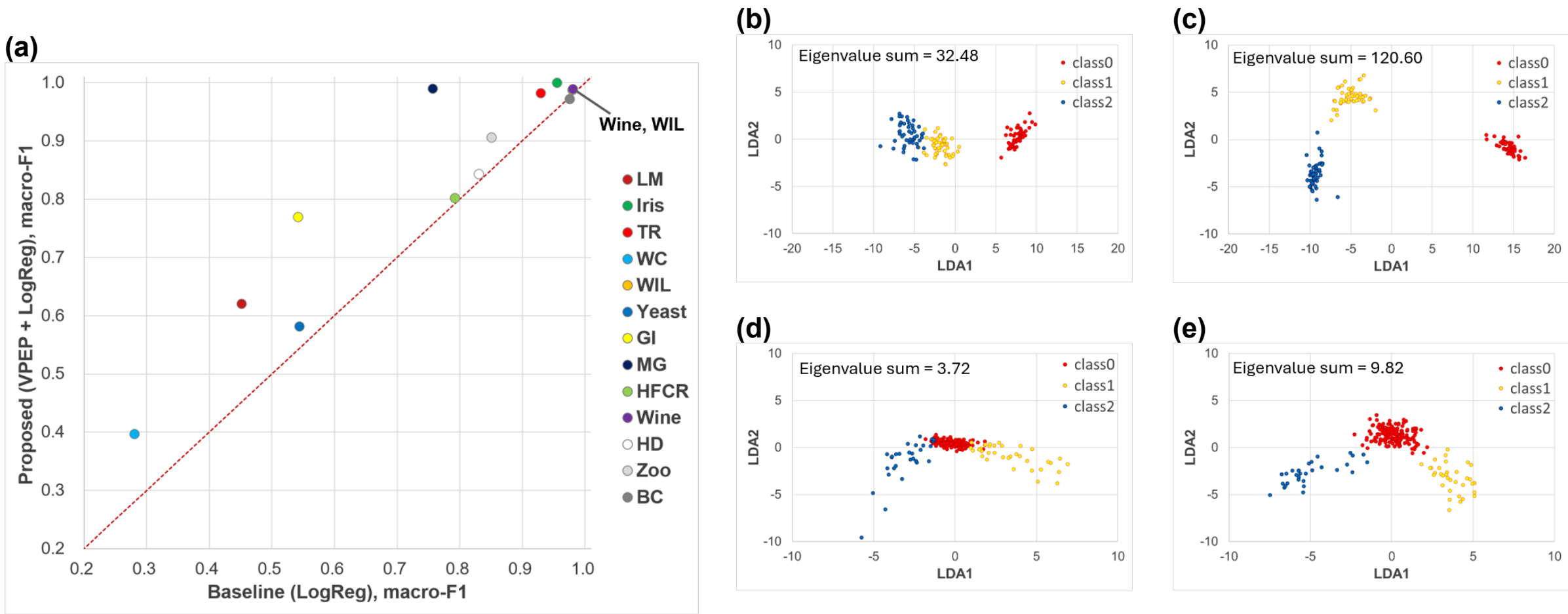

**FIGURE 5. Classification performance and representative feature-space visualization. (a) Parity plot of the mean macro-F1 scores for the baseline LogReg and VPEP + LogReg pipelines across 13 datasets; the dashed line indicates equal performance. LDA projections for (b) Iris baseline, (c) Iris with VPEP, (d) Thyroid (New) baseline, and (e) Thyroid (New) with VPEP. The baseline and VPEP projections are computed from $x$ and $z$, respectively. The Fisher eigenvalue sum is defined here as the sum of the non-negative eigenvalues of $S_W^{+}S_B$, where $S_W$ and $S_B$ denote the within-class and between-class scatter matrices, respectively, and $S_W^{+}$ denotes the Moore–Penrose pseudoinverse; larger values indicate greater class separability. LM: Land Mines-1, TR: Thyroid (New), WC: Wholesale Customers, WIL: Wireless Indoor Localization, GI: Glass Identification, MG: MAGIC Gamma Telescope, HFCR: Heart Failure Clinical Records, HD: Heart Disease (processed Cleveland), BC: Breast Cancer Wisconsin (Diagnostic).**

### 3) RESULTS: REGRESSION

Regression performance is evaluated by comparing the proposed VPEP + LinReg pipeline with the LinReg baseline using standardized input features without photonic encoding across 12 regression benchmarks derived from public datasets available through the UCI Machine Learning Repository [41] and scikit-learn [34]. The benchmarks span 4–16 input features and diverse target variables. The following benchmarks are used: Combined Cycle Power Plant (CCPP) [58], Airfoil Self-Noise (ASN) [59], Yacht Hydrodynamics (YH) [60], Real Estate Valuation (REV) [61], Energy Efficiency—heating/cooling load (EE-1/EE-2) [62], Concrete Compressive Strength (CCS) [63], Diabetes (DIAB) [64], Wine Quality—red/white (WQ-R/WQ-W) [65], [66], and Condition-Based Maintenance of Naval Propulsion Plants (CBM-1/CBM-2) [67].

Fig. 6(a) compares the mean $R^2$ values for the baseline and VPEP + LinReg pipelines in a parity plot. VPEP + LinReg achieves a higher mean $R^2$ than the baseline on 9 of the 12 benchmarks while retaining the same lightweight linear-regression back end. Particularly large improvements are observed for YH, CCS, ASN, and CBM-1, demonstrating that the fixed photonic transformation can substantially improve regression performance for some benchmarks. Performance decreases are observed for EE-1, EE-2, and most notably CBM-2. The pronounced decrease for CBM-2 has not yet been linked to a specific dataset characteristic or aspect of the photonic transformation. More generally, no simple dataset-level property has yet been identified that reliably predicts whether, or to what extent, VPEP encoding will improve regression performance. Full numerical results, including RMSE, $R^2$, and cross-validation statistics, are provided in Table S-II.

To further examine the regression behavior, Figs. 6(b)–(e) show representative predicted-versus-target plots for YH and CCS. Compared with the corresponding baseline results, the VPEP + LinReg predictions more closely follow the ideal $y_p$

= $y_t$ relationship. For YH, $R^2$ increases from 0.6250 to 0.9186 while RMSE decreases from 9.0882 to 4.2223. For CCS, $R^2$ increases from 0.6028 to 0.7517 while RMSE decreases from 10.4659 to 8.2681. These examples illustrate that the feature representation generated by the fixed VPEP front end can make some regression relationships substantially more amenable to the same lightweight linear back end.

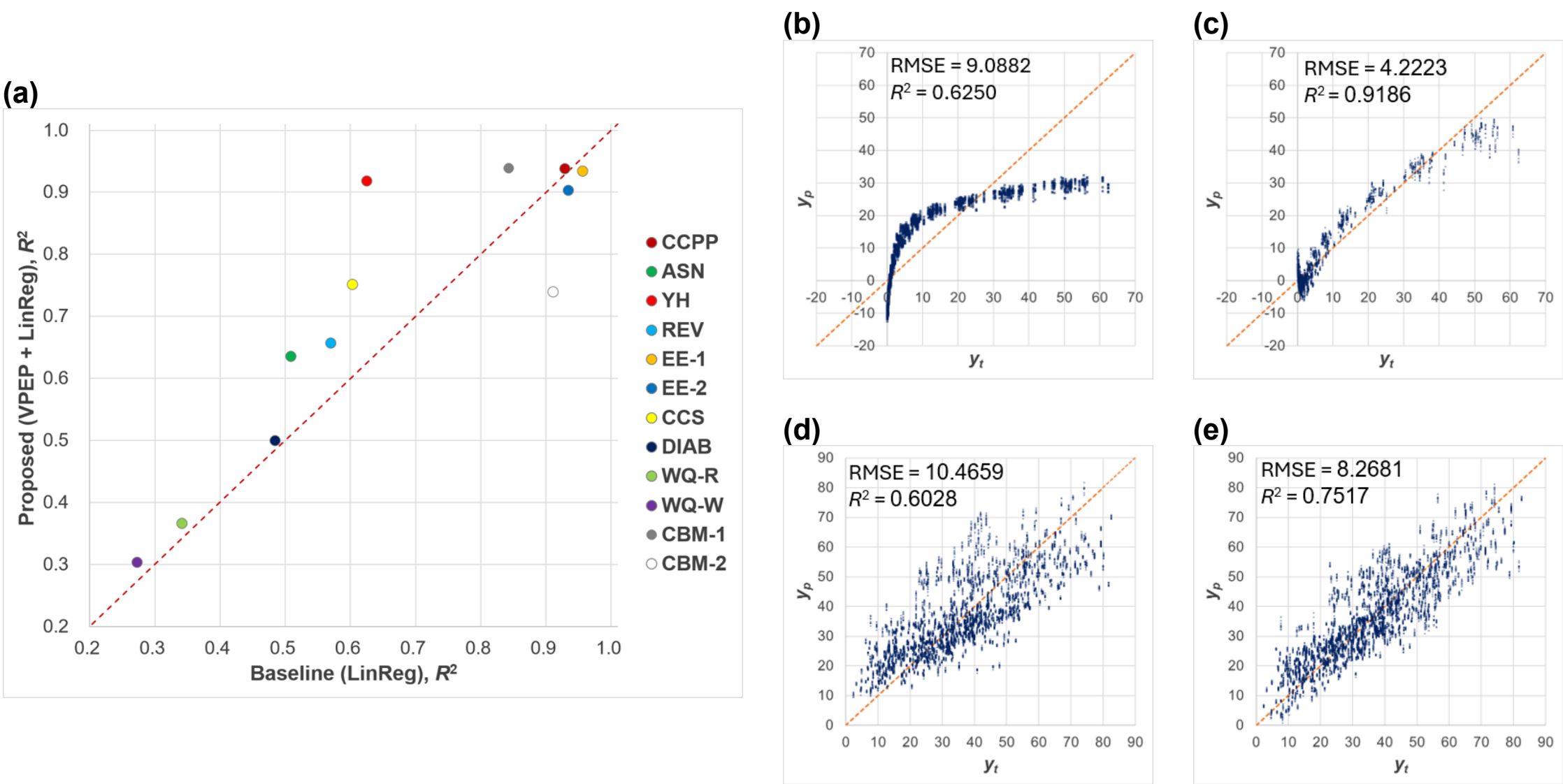

**FIGURE 6. Regression performance and representative predicted-versus-target plots. (a) Parity plot of the mean $R^2$ values for the baseline LinReg and VPEP + LinReg pipelines across 12 regression benchmarks; the dashed line indicates equal performance. Predicted-versus-target plots for (b) YH baseline, (c) YH with VPEP, (d) CCS baseline, and (e) CCS with VPEP. In (b)–(e), the dashed lines indicate perfect agreement ($y_p = y_t$), and the reported values show RMSE and $R^2$. CCPP: Combined Cycle Power Plant, ASN: Airfoil Self-Noise, YH: Yacht Hydrodynamics, REV: Real Estate Valuation, EE-1/EE-2: Energy Efficiency—heating/cooling load, CCS: Concrete Compressive Strength, DIAB: Diabetes, WQ-R/WQ-W: Wine Quality-red/white, CBM-1/CBM-2: Condition-Based Maintenance of Naval Propulsion Plants.**

### C. A/D CONVERSION

#### 1) EVALUATION PROTOCOL

Fig. 7 illustrates the evaluation pipeline for the VPEP-based A/D conversion task. A scalar input $x \in \{0, 1, ..., 1000\}$ is mapped to the 32-channel drive vector $\boldsymbol{u}$ and applied to the VPEP, yielding the measured 17-channel analog electrical output vector $\boldsymbol{z}$. Each channel of $\boldsymbol{z}$ is independently quantized into $L \in \{2, 3, 4\}$ levels, producing a 17-dimensional discrete code vector $\boldsymbol{b}$. The thresholds for each channel are determined from the measured distribution of $\boldsymbol{z}$ using a quantile-based rule to provide approximately equal occupancy among the quantization levels. Once determined, the thresholds are fixed during decoder calibration and evaluation. A digital decoder then reconstructs the input as $x'$ from the quantized vector $\boldsymbol{b}$.

The experimental conditions are summarized in Table II. We employ a lightweight decoder consisting of linear regression (LinReg) [36] followed by an optional one-dimensional lookup table (1D LUT) for residual correction, a widely used approach for the digital post-correction of ADC non-idealities [68], [69]. LinReg estimates $x$ from the 17-dimensional quantized vector $\boldsymbol{b}$ using least-squares fitting. When the LUT is enabled, it is indexed by the LinReg estimate and stores the mean reconstruction residual within each bin. The number of LUT bins is varied among 128, 256, 512, and 1024. When the LUT is not used, the decoder consists of LinReg alone. Because some LUT bins may contain no calibration samples, empty bins are filled via linear interpolation between neighboring non-empty bins. The decoder output is constrained to the calibrated input range of 0–1000. The replication factor $\rho$ and channel assignment (permutation $P$) are selected through preliminary evaluation and then held fixed throughout the reported A/D conversion experiments, with $\rho = 24$.

Unlike the classification and regression benchmarks, the A/D conversion task is treated as a calibration problem over a fixed and bounded operating range rather than as a generalization problem over unseen input samples. Accordingly, the full sweep of 1001 input points is used for threshold determination, decoder calibration, and metric evaluation, with no held-out test set. The reported results therefore characterize in-sample calibration performance over the specified operating range rather than out-of-sample generalization. The decoder is intentionally restricted to lightweight models to reflect implementation-efficient mixed-signal calibration and to avoid obscuring the characteristics of the photonic encoding and quantization stages with a highly expressive digital inverse model.

Performance is quantified using the normalized root-mean-square error (NRMSE), the 99th percentile of the absolute reconstruction error (p99 $|e|$), the monotonicity violation rate (MVR), and the quantization-noise-equivalent effective number of bits (ENOB), where $e = x' - x$. NRMSE is defined as the RMSE normalized by the full-scale input range, $x_{\max} - x_{\min}$. The p99 $|e|$ metric characterizes the upper tail of the absolute reconstruction error distribution. The MVR is defined as the fraction of adjacent input steps for which the reconstructed output decreases. By analogy with the ideal quantization noise relation underlying ENOB calculations [70]–[72], the quantization-noise-equivalent ENOB is calculated as

$$ENOB = \frac{1}{2} log_2 \left(\frac{1}{12\, NRMSE^2}\right), \qquad (7)$$

providing a bit-equivalent measure of the reconstruction resolution.

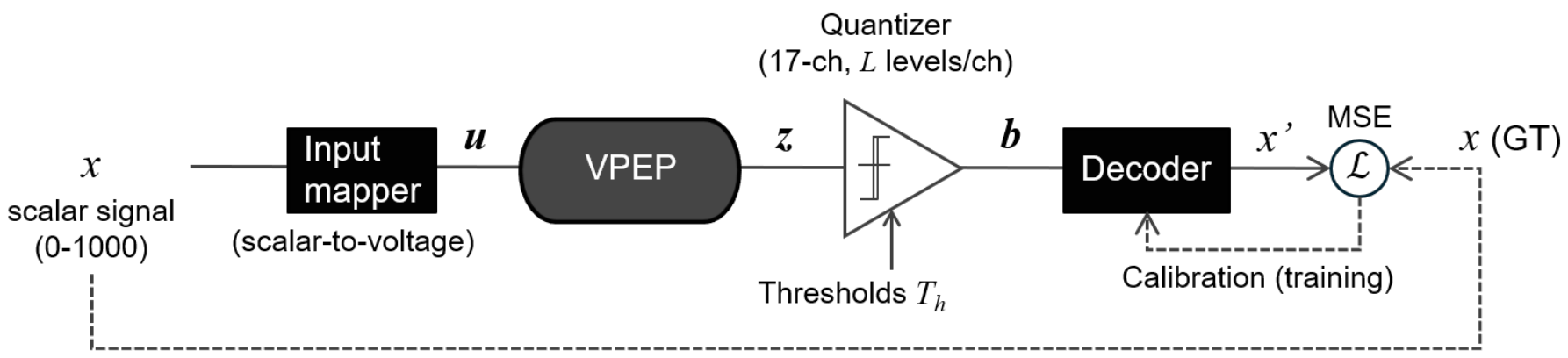

**FIGURE 7. Evaluation pipeline for the VPEP-based analog-to-digital conversion task. The pipeline comprises input mapping, VPEP encoding, per-channel quantile-based quantization, and lightweight digital decoding. MSE: mean squared error; GT: ground truth.**

TABLE II
EXPERIMENTAL CONDITIONS FOR ANALOG-TO-DIGITAL CONVERSION TASK

| | |
|---|---|
| **Decoder algorithm** | LinReg + 1D LUT |
| **LUT bins** | Not used / 128 / 256 / 512 / 1024 |
| **Calibration (training) objective** | MSE for reconstructing x from quantized vector b:<br>LinReg: least-squares fitting;<br>1D LUT: mean residual correction within each bin |
| **Input data $x$** | Scalar input sweep:<br>$x \in \{0,1,\ldots,1000\}$ (1001 points; unit step) |
| **Replication factor ρ** | 24; selected based on preliminary evaluation |
| **Channel assignment (permutation $P$ )** | Selected based on preliminary evaluation |
| **Quantizer** | Per-channel quantization:<br>$L$ = 2,3,4 levels/channel (17 channels) |
| **Thresholds $T_h$** | Channel-wise quantile-based thresholds;<br>fixed after calibration |
| **Calibration/evaluation protocol** | Full 1001-point sweep used for threshold setting, decoder calibration (training), and metric evaluation;<br>no held-out test set |
| **Performance metrics** | NRMSE, p99 $|e|$, MVR, ENOB |

LinReg: linear regression; LUT: lookup table; MSE: mean squared error; NRMSE: normalized root-mean-square error; MVR: monotonicity violation rate; ENOB: effective number of bits.

### 2) RESULTS

Fig. 8 summarizes the A/D conversion performance of the VPEP-based pipeline under the experimental conditions described in Table II. Detailed numerical results for the evaluated combinations of quantization level $L$ and decoder configuration are provided in Table S-III.

Fig. 8(a) shows the NRMSE, expressed as a percentage of full scale (%FS), as a function of the number of 1D LUT bins for $L$ = 2, 3, and 4. The leftmost point for each $L$ corresponds to LinReg without LUT correction. The inclusion of the 1D LUT consistently reduces the reconstruction error for all three quantization levels, with the NRMSE decreasing further as the number of LUT bins increases from 128 to 1024. This systematic improvement indicates that a substantial component of the residual error after linear decoding can be effectively compensated using a simple 1D correction indexed by the LinReg output.

Among the evaluated configurations, $L = 3$ with LinReg + 1D LUT using 1024 bins achieves the lowest NRMSE, approximately 0.75%FS. Notably, increasing the quantization level from $L = 3$ to L = 4 does not further improve the end-to-end reconstruction accuracy under the present decoder configuration. Although $L = 4$ provides a larger code space and greater code diversity, its reconstruction error remains higher than that obtained with $L = 3$. This result indicates that increasing the per-channel quantization level does not necessarily translate directly into improved end-to-end reconstruction when the decoder complexity is constrained.

Fig. 8(b) shows the decoded output $x'$ as a function of the input $x$ for the best evaluated configuration. The decoded output closely follows the ideal relation $x' = x$ over the calibrated input range, achieving an NRMSE of approximately 0.75%FS and a quantization-noise-equivalent ENOB of 5.27 bits. Fig. 8(c) shows the corresponding residual $e = x' - x$, with a 99th-percentile absolute reconstruction error of p99 $|e|$ = 23 LSB. Here, LSB denotes one input step (1/1000 FS). The MVR is 6.1%, indicating that local non-monotonic steps remain despite the low overall reconstruction error. These results demonstrate high reconstruction fidelity over the calibrated range while also identifying monotonicity as an important target for further improvement.

Overall, the results demonstrate that the same fixed VPEP front end used in the preceding classification and regression experiments can also be extended to mixed-signal conversion by combining its 17-channel encoded outputs with per-channel quantization and lightweight digital decoding. Under the best evaluated condition, the scalar input is reconstructed over the calibrated operating range with an NRMSE of approximately 0.75%FS using only LinReg and a 1D LUT for digital correction. This result demonstrates the feasibility of extending the common fixed photonic encoding hardware to A/D conversion without relying on a highly expressive digital inverse model.

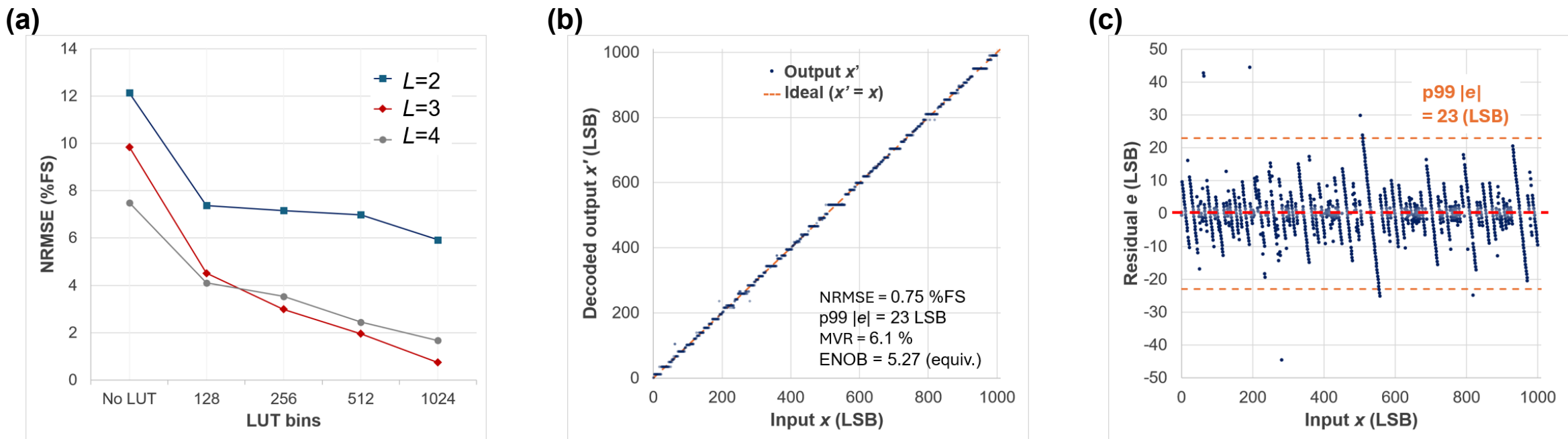

FIGURE 8. **Analog-to-digital conversion performance of the VPEP-based pipeline. (a) Normalized root-mean-square error (NRMSE) (% of full scale, FS) versus 1D lookup table (LUT) size for $L$ = 2, 3, and 4 quantization levels; the leftmost points indicate LinReg without LUT correction. (b) Decoded output $x'$ versus input $x$ for the best evaluated configuration ($L$ = 3, LinReg + 1D LUT with 1024 bins); the dashed line indicates the ideal relation $x' = x$. (c) Residual $e = x' - x$ for the same configuration; the outer dashed lines indicate ±p99 |e| = ±23 (LSB). One LSB corresponds to one input step (1/1000 FS). MVR: monotonicity violation rate, ENOB: effective number of bits, p99 |e|: 99th percentile of absolute reconstruction error.**

### D. PHOTONIC INFORMATION ENCODING AND RECOVERY

#### 1) EVALUATION PROTOCOL

Fig. 9 illustrates the evaluation pipeline for photonic information encoding and recovery using VPEP. In the envisioned in-vehicle scenario, VPEP transforms digital information into a device-dependent encoded representation, which is subsequently decoded at a remote node, such as a central computing unit. The input $\boldsymbol{x}$ is a 32-bit vector formed from four ASCII characters (4 bytes). The input mapper converts $\boldsymbol{x}$ into the 32-channel drive vector $\boldsymbol{u}$, which is applied to VPEP to generate the 17-channel encoded output vector $\boldsymbol{z}$. A receiver-side decoder is then trained in a supervised manner to recover the original information from $\boldsymbol{z}$, yielding the reconstructed four-character block $\boldsymbol{x}'$.

The experimental conditions are summarized in Table III. Three receiver-side decoders are evaluated: LDA [56], LogReg, and a multilayer perceptron (MLP) with one hidden layer of 128 units. All three decoders are implemented using scikit-learn [34] with the LinearDiscriminantAnalysis [57], LogisticRegression [35], and MLPClassifier [73] classes, respectively. Information recovery is formulated as byte-wise supervised multiclass classification. Each of the four character positions is decoded independently as a C-class classification problem, and the four predicted characters are concatenated to reconstruct the original block. LDA and LogReg provide lightweight linear decoders, whereas the shallow MLP provides a simple nonlinear decoder for evaluating the benefit of modest nonlinear processing in recovering the VPEP-encoded information.

The symbol set size is varied as $C \in \{10, 26, 32, 52, 94\}$, corresponding to digits (e.g., 0, 1, 2), lowercase letters (e.g., a, b, c), special characters (e.g., #, *, @), uppercase and lowercase letters (e.g., A, B, a, b), and printable ASCII characters excluding space (e.g., A, 1, #), respectively. The dataset contains 10,000 four-character blocks for $C$ = 10 and 200,000 blocks for each of $C$ = 26, 32, 52, and 94. For $C$ = 10, all $C^4 = 10^4$ possible four-character blocks are exhaustively enumerated. For the larger symbol sets, where the number of possible blocks grows as $C^4$, 200,000 blocks are generated by sampling characters from the corresponding symbol set.

Performance is evaluated using repeated five-fold cross-validation with five repeats, implemented with RepeatedKFold [38], treating each four-character block as one sample. The recovery performance is quantified using character error rate (CER) and four-character block error rate (BLER). The CER measures the fraction of incorrectly recovered characters, whereas BLER counts a block as erroneous if at least one of its four characters is incorrectly recovered.

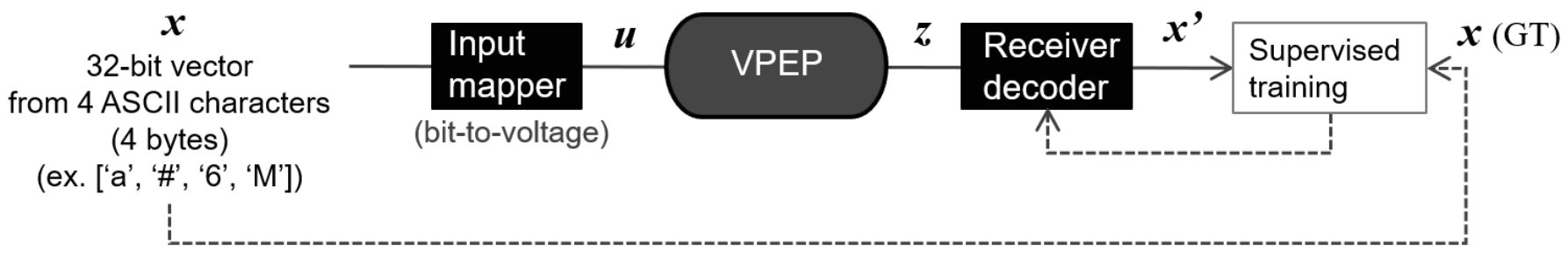

**FIGURE 9. Evaluation pipeline for photonic information encoding and recovery. The pipeline comprises input mapping, VPEP encoding, and supervised receiver-side decoding. GT: ground truth.**

TABLE III
EXPERIMENTAL CONDITIONS FOR PHOTONIC INFORMATION ENCODING AND RECOVERY

| | |
|---|---|
| **Decoder algorithm** | LDA, LogReg, MLP (1 hidden layer, 128 units) |
| **Decoding formulation** | Byte-wise multiclass classification |
| **Input data *x*** | 32-bit vector from 4 ASCII characters (4 bytes) |
| **Symbol set size *C*** | 10 (digits), 26 (lowercase), 32 (special characters), 52 (uppercase + lowercase), 94 (printable ASCII, excluding space) |
| **Dataset size (blocks)** | 10k ($C$ = 10), 200k ($C$ = 26, 32, 52, 94) |
| **Evaluation protocol** | Repeated $k$-fold CV |
| **Folds / repeats** | $k$ = 5, repeats = 5 |
| **Performance metrics** | CER, BLER |

LDA: linear discriminant analysis; LogReg: logistic regression; MLP: multilayer perceptron; ASCII: American Standard Code for Information Interchange; CV: cross-validation; CER: character error rate; BLER: block error rate.

### 2) RESULTS

Fig. 10 summarizes the information recovery performance in the VPEP-based photonic encoding experiment under the conditions described in Table III. Detailed numerical results, including the mean and standard deviation of the CER and BLER over repeated cross-validation, are provided in Table S-IV.

Fig. 10(a) shows the character reconstruction accuracy as a function of the symbol set size $C$ for the byte-wise LDA, LogReg, and MLP decoders. For the smallest symbol set ($C$ = 10, digits), all three decoders achieve high character accuracy, with the MLP providing the highest performance (CER = 1.96%, corresponding to approximately 98% character accuracy). As $C$ increases, character accuracy decreases for all three decoders, but the decrease is substantially smaller for the nonlinear MLP. For example, at $C$ = 26 (lowercase letters), the MLP achieves a CER of 6.41% (approximately 94% character accuracy), whereas the linear decoders exhibit CER values of approximately 29–34%. The performance advantage of the MLP remains pronounced for the larger symbol sets. At $C$ = 94, the MLP achieves a CER of 52.07% and a BLER of 91.88%, outperforming the linear decoders but indicating that recovery becomes increasingly challenging as the information space expands. The consistently higher performance of the MLP demonstrates that modest nonlinear receiver-side processing can substantially improve recovery from the VPEP-encoded representation, particularly for larger symbol sets.

The block-level results further illustrate the increased difficulty of multi-character recovery as the symbol set expands. Because a four-character block is counted as correct only when all four characters are recovered correctly, increases in CER lead to a correspondingly rapid increase in BLER. Thus, even when the MLP maintains substantially higher character accuracy than that of the linear decoders, recovery of complete four-character blocks becomes more challenging for larger symbol sets.

Fig. 10(b) further resolves the MLP performance by byte position. The four byte positions (byte0–byte3) correspond to the four ASCII characters in the 32-bit input. These positions exhibit similar character accuracies for $C$ = 10, whereas noticeable differences emerge among byte positions for larger symbol sets. To examine this position dependence in greater detail, Fig. 10(c) shows an additional diagnostic bit-wise MLP analysis at $C$ = 94. The bit error rate (BER) varies across the active input-bit positions, indicating position-dependent differences in information recovery. These differences may reflect both the mapping of input bits to the VPEP modulation channels and the statistical occurrence and predictability of individual bits in the ASCII representation. Bits 0, 8, 16, and 24 correspond to constant most-significant bits and therefore exhibit zero BER. These observations suggest that improved input mapping and channel assignment may provide a route toward reducing position-dependent recovery errors.

Overall, the results demonstrate that the device-dependent representation generated by the fixed VPEP front end retains sufficient information for receiver-side recovery of multi-character digital inputs using lightweight electronic decoders. The clear advantage of the one-hidden-layer MLP over the linear decoders further shows that modest nonlinear processing can substantially extend the recoverable information space without requiring a highly complex back end. Together with the classification, regression, and A/D conversion results, this experiment shows that the same fixed VPEP hardware can operate across machine learning, mixed-signal conversion, and information encoding and recovery, further illustrating its cross-application versatility.

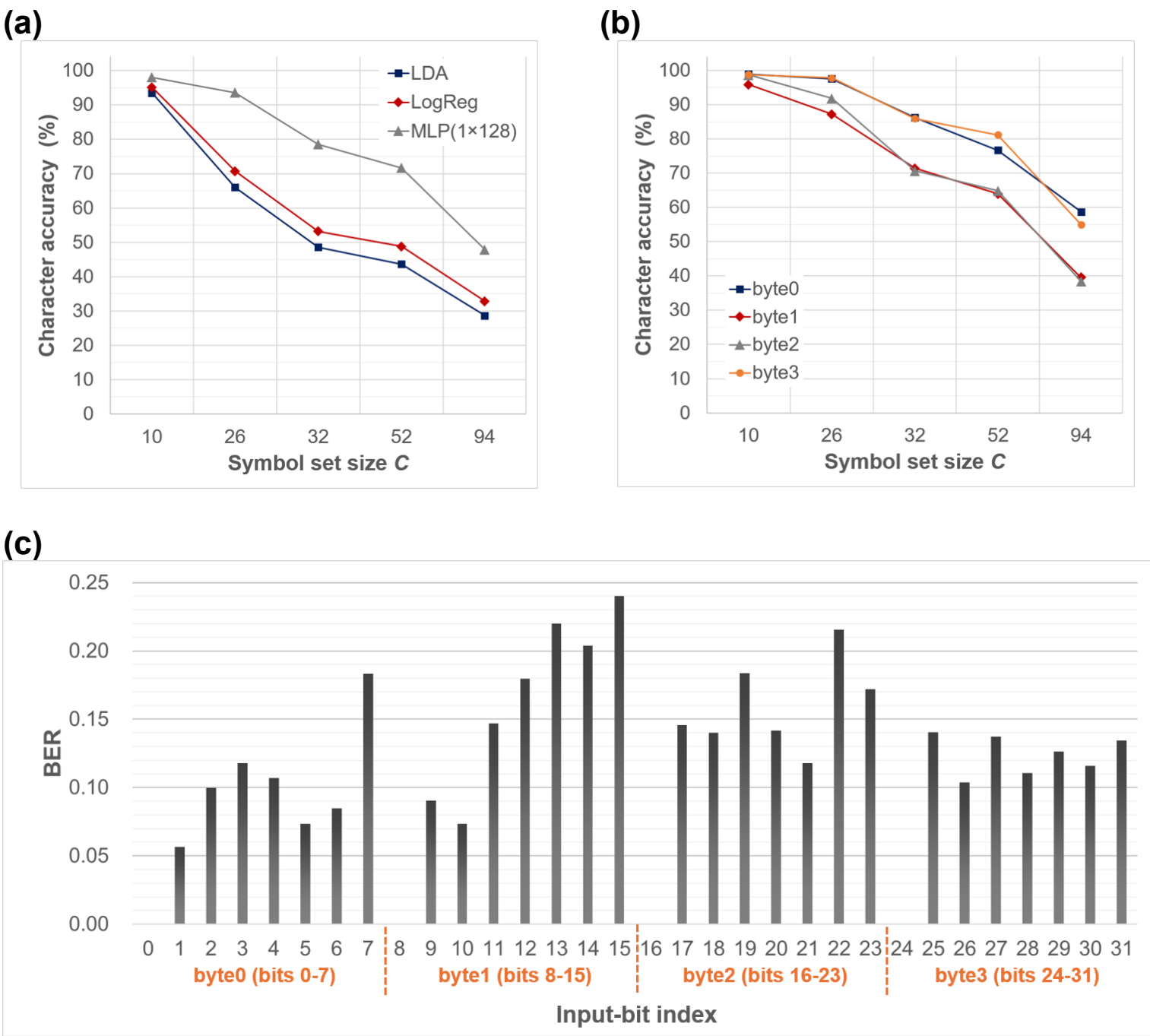

**FIGURE 10. Photonic information encoding and recovery performance obtained using VPEP. (a) Mean character accuracy versus symbol set size *C* for the byte-wise LDA, LogReg, and MLP decoders. (b) Mean character accuracy by byte position obtained using the byte-wise MLP decoder. (c) Mean bit error rate (BER) for each input bit from an additional diagnostic bit-wise MLP analysis at *C* = 94, grouped by byte position. Bits 0, 8, 16, and 24 correspond to constant most-significant bits in the ASCII representation and therefore exhibit zero BER.**

## IV. DISCUSSION

This work investigated whether a single, fixed silicon-photonic encoding core can serve as a common front end for fundamentally different edge-processing functions. The experimental results demonstrate the feasibility of this concept. Without task-specific reconfiguration of the photonic encoding core, the same fabricated VPEP chip was applied to classification, regression, A/D conversion, and photonic information encoding and recovery. The significance of these results lies not in achieving state-of-the-art performance for any individual task, but in demonstrating cross-application versatility with one unchanged photonic core—a property well aligned with the emphasis on hardware commonality and implementation simplicity in automotive edge systems.

The classification and regression experiments provide the most direct assessment of VPEP as a feature transformation front end. For classification, VPEP + LogReg achieved a higher mean macro-F1 score than the LogReg baseline on 12 of the 13 evaluated datasets, while representative LDA analyses indicated improved linear class separability after photonic encoding. For regression, VPEP + LinReg achieved a higher mean $R^2$ than the LinReg baseline on 9 of the 12 evaluated benchmarks. These results indicate that the nonlinear representation generated by phase modulation, passive optical mixing, and photodetection can reshape the input feature space into a form that is more amenable to lightweight linear back ends. The benefit is dataset-dependent, however, and performance degradation was observed for a small number of classification and regression benchmarks. No simple dataset-level property has yet been identified that reliably predicts whether, or to what extent, VPEP encoding will improve task performance.

The A/D conversion and information encoding experiments extend the evaluation beyond conventional machine learning tasks. In the A/D conversion pipeline, VPEP encoding combined with per-channel quantization and lightweight digital decoding reconstructed the scalar input over the calibrated operating range, with the best evaluated configuration achieving an in-sample NRMSE of approximately 0.75%FS. A 1D LUT substantially improved reconstruction over LinReg alone, showing that residual errors can be effectively compensated without relying on a highly expressive inverse model. Increasing the per-channel quantization level did not monotonically improve reconstruction accuracy, indicating that greater code diversity does not necessarily translate directly into better end-to-end performance when decoder complexity is constrained. In the information encoding experiment, receiver-side recovery became increasingly difficult as the symbol set size increased, while the nonlinear MLP decoder consistently outperformed the linear decoders for larger symbol sets. The observed position-dependent errors may reflect the input mapping, the statistical properties of the ASCII representation, and the physical encoding characteristics of the device. Improved mapping, redundancy, and lightweight error-control processing may therefore provide practical routes toward more reliable recovery within the same fixed-core architecture.

The information encoding results are also relevant beyond direct symbol recovery. Because the 32-bit digital input is transformed into a lower-dimensional, device-dependent 17-channel encoded representation, the same physical transformation may provide a basis for future applications involving dimensionality reduction, compact information representation, data obfuscation, and physical-layer security. These directions are particularly interesting because they exploit the fixed device-defined transformation itself, rather than requiring application-specific reconfiguration of the photonic core.

Taken together, the four evaluations clarify what versatility means for VPEP: the photonic encoding core remains fixed, while application-dependent adaptation is handled at its periphery. The input mapper determines how information is distributed across the 32 modulation channels, and the electronic back end interprets the resulting 17-channel encoded representation according to the target function. Consequently, the replication factor and channel assignment can influence performance even though the physical mixing matrix remains unchanged. Equally important, the required peripheral processing remained relatively lightweight. Linear models were sufficient to improve performance for most of the evaluated classification datasets and regression benchmarks, a 1D LUT effectively improved A/D conversion reconstruction, and a shallow MLP substantially improved information recovery when linear decoders became insufficient. These results show that useful application adaptation does not necessarily require a highly expressive digital back end. This division of functionality—a common fixed photonic encoding core providing the physical transformation and lightweight peripheral processing adapting the encoded representation to each target function—embodies the central concept of VPEP.

This architecture also highlights an important open problem in input mapping. In the present experiments, mapper parameters were selected through preliminary evaluation, and a general criterion for selecting them for a new dataset or task has not yet been established. Developing a systematic and efficient mapping strategy that adapts new input representations to a fixed photonic core without extensive task-by-task search will therefore be important for broader deployment. In particular, clarifying the relationships among input statistics, mapper configuration, the fixed photonic transformation, and downstream performance may enable more efficient mapper selection and provide deeper insight into when photonic encoding is beneficial.

Several challenges remain before VPEP can be adopted as a practical automotive edge platform. First, the present device has fixed physical input and output dimensions of 32 and 17 channels, respectively. Simply increasing the number of physical ports would increase the device area, optical routing complexity and loss, electrical I/O requirements, and packaging complexity. Future scaling should therefore consider approaches that increase the effective encoding dimension without a proportional increase in physical I/O, such as wavelength- or time-domain multiplexing or the coordinated use of multiple compact encoding cores.

Second, thermal stability is a particularly important challenge for silicon-photonic implementations because the refractive index of silicon is temperature-dependent, causing temperature variations to alter the optical phase and interference conditions throughout the photonic circuit [21], [22]. Accordingly, the VPEP chip was operated at a controlled temperature of 35 °C in the present experiments. In addition, because the present prototype employs thermo-optic PMs, local heating and associated thermal crosstalk may further perturb the phase distribution across the chip. The temperature dependence, long-term drift, operational repeatability, and device-to-device variation have not yet been systematically characterized.

For automotive deployment, these considerations extend to long-term reliability under environmental stresses representative of the intended installation location. The fixed photonic transformation must remain sufficiently stable over the required service lifetime under wide-range temperature operation, temperature cycling, humidity, vibration, and mechanical shock. This requirement is particularly important for VPEP because the input mapping and electronic back ends are adapted to the device-defined transformation; long-term drift or environmentally induced changes in this transformation could degrade application performance or necessitate recalibration. Automotive integrated circuits and optoelectronic semiconductors are commonly qualified using frameworks such as AEC-Q100 [74] and AEC-Q102 [75], respectively; however, an integrated photonic–electronic platform such as VPEP must additionally consider the combined reliability of photonic circuitry, optical coupling, and packaging. Future work should therefore establish appropriate automotive mission profiles and reliability criteria for VPEP and evaluate whether compensation or recalibration can maintain the required performance over the service lifetime.

Third, the potential energy and latency advantages of photonic processing have not yet been verified at the system level. Although passive propagation and interference can perform optical mixing with high bandwidth and without sequential electronic arithmetic [15], [16], overall energy and latency will also depend on the PMs, optical source, PDs, data converters, control electronics, and other electrical–optical interface overheads. The thermo-optic PMs used in the present prototype, in particular, were not optimized for low-power operation. Quantifying system-level energy per transformation and end-to-end latency and investigating more energy-efficient modulation and interface technologies will be essential for determining how effectively the potential bandwidth and passive-processing advantages of the photonic core translate into system-level benefits.

Finally, the present application evaluations represent proof-of-concept studies rather than deployment-complete implementations. The classification and regression benchmarks were selected to examine feasibility across diverse tabular datasets rather than to reproduce the full complexity of automotive workloads. The A/D conversion experiment was evaluated over a bounded input range using full-sweep calibration. The information encoding experiment demonstrated recovery from a device-dependent representation. However, the feasibility of security-related applications, such as data obfuscation and physical-layer security, remains to be evaluated under appropriate system and threat models. More broadly, future studies should extend VPEP to more deployment-oriented automotive tasks, such as sensor diagnostics, condition monitoring, anomaly detection, and compact perception or signal processing functions, under realistic operating constraints. Ultimately, direct comparisons with specialized electronic and programmable photonic implementations will be necessary to identify the conditions under which a common fixed photonic front end provides meaningful system-level advantages.

## V. CONCLUSION

A single fabricated VPEP chip, with its photonic encoding core left unchanged, was experimentally demonstrated across four distinct processing functions: classification, regression, A/D conversion, and photonic information encoding and recovery. By keeping the photonic core common and shifting application adaptation to input mapping and lightweight peripheral processing, rather than tailoring the photonic hardware to each individual function, VPEP provides a system-level design approach in which application versatility is achieved without reconfiguring the photonic core. This approach is complementary to reconfigurable photonic architectures.

The results demonstrate the cross-application versatility of the same fixed photonic hardware across machine learning, mixed-signal conversion, and information encoding and recovery. With further advances in scalable encoding, low-power interfaces, efficient task adaptation, and robust system integration, this approach may provide a practical path toward versatile photonic front ends for future automotive edge systems and other distributed sensing and computing platforms.

## ACKNOWLEDGMENT

The authors thank Tatsuya Yamashita and Yukihiro Tadokoro for their support and encouragement in conducting this research. The authors acknowledge the UCI Machine Learning Repository and scikit-learn for providing the public datasets used in the classification and regression experiments. The authors thank FORTE Science Communications for English language editing. The authors also acknowledge the use of ChatGPT (OpenAI) and Copilot (Microsoft) during the preparation of this manuscript to assist with Japanese-to-English translation and bibliographic checks. All AI-assisted content was subsequently reviewed and edited by the authors, who take full responsibility for the final content of the manuscript.

## REFERENCES

[1] E. Yurtsever, J. Lambert, A. Carballo, and K. Takeda, "A survey of autonomous driving: Common practices and emerging technologies," IEEE Access, vol. 8, pp. 58443–58469, 2020, doi: 10.1109/ACCESS.2020.2983149.

[2] Z. Wang, Y. Wu, and Q. Niu, "Multi-sensor fusion in automated driving: A survey," IEEE Access, vol. 8, pp. 2847–2868, 2020, doi: 10.1109/ACCESS.2019.2962554.

[3] V. Bandur, G. Selim, V. Pantelic, and M. Lawford, "Making the case for centralized automotive E/E architectures," IEEE Trans. Veh. Technol., vol. 70, no. 2, pp. 1230–1245, Feb. 2021, doi: 10.1109/TVT.2021.3054934.

[4] H. Askaripoor, M. Hashemi Farzaneh, and A. Knoll, "E/E architecture synthesis: Challenges and technologies," Electronics, vol. 11, no. 4, Art. no. 518, Feb. 2022, doi: 10.3390/electronics11040518.

[5] L. Mauser, H. Patil, L. Patino Studencki, and S. Wagner, "ConvIO: A framework for convergence of inputs and outputs in automotive zonal E/E architectures," IEEE Access, vol. 14, pp. 98616–98632, 2026, doi: 10.1109/ACCESS.2026.3707952.

[6] W. Shi, J. Cao, Q. Zhang, Y. Li, and L. Xu, "Edge computing: Vision and challenges," IEEE Internet Things J., vol. 3, no. 5, pp. 637–646, Oct. 2016, doi: 10.1109/JIOT.2016.2579198.

[7] S. Liu, L. Liu, J. Tang, B. Yu, Y. Wang, and W. Shi, "Edge computing for autonomous driving: Opportunities and challenges," Proc. IEEE, vol. 107, no. 8, pp. 1697–1716, Aug. 2019, doi: 10.1109/JPROC.2019.2915983.

[8] S.-C. Lin, K.-C. Chen, and A. Karimoddini, "SDVEC: Software-defined vehicular edge computing with ultra-low latency," IEEE Commun. Mag., vol. 59, no. 12, pp. 66–72, Dec. 2021, doi: 10.1109/MCOM.004.2001124.
[9] M. Di Natale and A. L. Sangiovanni-Vincentelli, "Moving from federated to integrated architectures in automotive: The role of standards, methods and tools," Proc. IEEE, vol. 98, no. 4, pp. 603–620, Apr. 2010, doi: 10.1109/JPROC.2009.2039550.
[10] L. Braun, M. Armbruster, and E. Sax, "Stakeholder issues concerning the automotive E/E-architecture," in Proc. Int. Conf. Electr. Syst. Aircr., Railway, Ship Propuls. Road Veh. Int. Transp. Electrific. Conf. (ESARS-ITEC), Toulouse, France, Nov. 2016, pp. 1–6, doi: 10.1109/ESARS-ITEC.2016.7841340.
[11] R. P. Dick et al., "Research challenges for energy-efficient computing in automated vehicles," Computer, vol. 56, no. 3, pp. 47–58, Mar. 2023, doi: 10.1109/MC.2022.3180987.
[12] D. Katare, D. Perino, J. Nurmi, M. Warnier, M. Janssen, and A. Y. Ding, "A survey on approximate edge AI for energy efficient autonomous driving services," IEEE Commun. Surveys Tuts., vol. 25, no. 4, pp. 2714–2754, 2023, doi: 10.1109/COMST.2023.3302474.
[13] X. Chen et al., "The emergence of silicon photonics as a flexible technology platform," Proc. IEEE, vol. 106, no. 12, pp. 2101–2116, Dec. 2018, doi: 10.1109/JPROC.2018.2854372.
[14] X. Zhou, D. Yi, D. W. U. Chan, and H. K. Tsang, "Silicon photonics for high-speed communications and photonic signal processing," npj Nanophoton., vol. 1, Art. no. 27, Jul. 2024, doi: 10.1038/s44310-024-00024-7.
[15] D. A. B. Miller, "Attojoule optoelectronics for low-energy information processing and communications," J. Lightwave Technol., vol. 35, no. 3, pp. 346–396, Feb. 2017, doi: 10.1109/JLT.2017.2647779.
[16] B. J. Shastri et al., "Photonics for artificial intelligence and neuromorphic computing," Nat. Photonics, vol. 15, no. 2, pp. 102–114, Feb. 2021, doi: 10.1038/s41566-020-00754-y.
[17] Y. Shen et al., "Deep learning with coherent nanophotonic circuits," Nat. Photonics, vol. 11, no. 7, pp. 441–446, Jul. 2017, doi: 10.1038/nphoton.2017.93.
[18] J. Feldmann et al., "Parallel convolutional processing using an integrated photonic tensor core," Nature, vol. 589, no. 7840, pp. 52–58, Jan. 2021, doi: 10.1038/s41586-020-03070-1.
[19] X. Xu et al., "11 TOPS photonic convolutional accelerator for optical neural networks," Nature, vol. 589, no. 7840, pp. 44–51, Jan. 2021, doi: 10.1038/s41586-020-03063-0.
[20] D. Pérez et al., "Multipurpose silicon photonics signal processor core," Nat. Commun., vol. 8, Art. no. 636, Sep. 2017, doi: 10.1038/s41467-017-00714-1.
[21] W. Bogaerts et al., "Programmable photonic circuits," Nature, vol. 586, pp. 207–216, Oct. 2020, doi: 10.1038/s41586-020-2764-0.
[22] X. Xu et al., "Self-calibrating programmable photonic integrated circuits," Nat. Photonics, vol. 16, no. 8, pp. 595–602, Aug. 2022, doi: 10.1038/s41566-022-01020-z.
[23] H. Feng and C. Wang, "Breaking the power wall in programmable photonics," Nat. Photonics, vol. 20, no. 8, pp. 861–862, Aug. 2026, doi: 10.1038/s41566-026-01960-w.
[24] K. Vandoorne et al., "Experimental demonstration of reservoir computing on a silicon photonics chip," Nat. Commun., vol. 5, Art. no. 3541, Mar. 2014, doi: 10.1038/ncomms4541.
[25] S. Ortín, M. C. Soriano, L. Pesquera, D. Brunner, D. San-Martín, I. Fischer, C. R. Mirasso, and J. M. Gutiérrez, "A unified framework for reservoir computing and extreme learning machines based on a single time-delayed neuron," Sci. Rep., vol. 5, Art. no. 14945, Oct. 2015, doi: 10.1038/srep14945.
[26] S. Biasi, R. Franchi, L. Cerini, and L. Pavesi, "An array of microresonators as a photonic extreme learning machine," APL Photon., vol. 8, no. 9, Art. no. 096105, Sep. 2023, doi: 10.1063/5.0156189.
[27] L. Huang and J. Yao, "Multi-task photonic time-delay reservoir computing based on polarization modulation," Opt. Lett., vol. 47, no. 24, pp. 6464–6467, Dec. 2022, doi: 10.1364/OL.479472.
[28] B. J. Giron Castro, C. Peucheret, D. Zibar, and F. Da Ros, "Multi-task photonic reservoir computing: Wavelength division multiplexing for parallel computing with a silicon microring resonator," Adv. Opt. Technol., vol. 13, Art. no. 1471239, Oct. 2024, doi: 10.3389/aot.2024.1471239.
[29] A. Mizuno, I. Takai, M. Nakai, A. Miyamichi, R. Konishi, and S. Sunada, "General-purpose nonlinear function approximation via linear integrated photonics," 2026, arXiv:2606.23159, doi: 10.48550/arXiv.2606.23159.
[30] R. Qin et al., "Towards direct sampling of ultra-broadband microwave signals: A review of integrated photonic analog-to-digital converter technologies," J. Lightwave Technol., vol. 43, no. 21, pp. 9793–9811, Nov. 2025, doi: 10.1109/JLT.2025.3593265.
[31] S. Yang and H. Jia, "Reconfigurable integrated silicon photonic analog-to-digital converter," J. Lightwave Technol., vol. 44, no. 8, pp. 2996–3003, Apr. 2026, doi: 10.1109/JLT.2026.3655727.
[32] F. Futami, K. Tanizawa, and K. Kato, "Y-00 quantum-noise randomized stream cipher using intensity modulation signals for physical layer security of optical communications," J. Lightwave Technol., vol. 38, no. 10, pp. 2774–2781, May 2020, doi: 10.1109/JLT.2020.2985709.
[33] F. Bin Tarik, A. Famili, Y. Lao, and J. D. Ryckman, "Scalable and CMOS compatible silicon photonic physical unclonable functions for supply chain assurance," Sci. Rep., vol. 12, Art. no. 15653, Sep. 2022, doi: 10.1038/s41598-022-19796-z.
[34] F. Pedregosa et al., "Scikit-learn: Machine learning in Python," J. Mach. Learn. Res., vol. 12, pp. 2825–2830, Oct. 2011.
[35] scikit-learn developers. "LogisticRegression." scikit-learn documentation. Accessed: May 8, 2026. [Online]. Available: https://scikit-learn.org/stable/modules/generated/sklearn.linear_model.LogisticRegression.html
[36] scikit-learn developers. "LinearRegression." scikit-learn documentation. Accessed: May 8, 2026. [Online]. Available: https://scikit-learn.org/stable/modules/generated/sklearn.linear_model.LinearRegression.html
[37] scikit-learn developers. "RepeatedStratifiedKFold." scikit-learn documentation. Accessed: May 8, 2026. [Online]. Available: https://scikit-learn.org/stable/modules/generated/sklearn.model_selection.RepeatedStratifiedKFold.html
[38] scikit-learn developers. "RepeatedKFold." scikit-learn documentation. Accessed: May 8, 2026. [Online]. Available: https://scikit-learn.org/stable/modules/generated/sklearn.model_selection.RepeatedKFold.html
[39] scikit-learn developers. "f1_score." scikit-learn documentation. Accessed: May 8, 2026. [Online]. Available: https://scikit-learn.org/stable/modules/generated/sklearn.metrics.f1_score.html
[40] scikit-learn developers. "r2_score." scikit-learn documentation. Accessed: May 8, 2026. [Online]. Available: https://scikit-learn.org/stable/modules/generated/sklearn.metrics.r2_score.html
[41] M. Kelly, R. Longjohn, and K. Nottingham. "The UCI Machine Learning Repository." Accessed: May 8, 2026. [Online]. Available: https://archive.ics.uci.edu/
[42] H. Kahraman, "Land Mines," UCI Machine Learning Repository, 2018, doi: 10.24432/C54C8Z.
[43] C. Yilmaz, H. T. Kahraman, and S. Söyler, "Passive mine detection and classification method based on hybrid model," IEEE Access, vol. 6, pp. 47870–47888, 2018, doi: 10.1109/ACCESS.2018.2866538.
[44] R. A. Fisher, "Iris," UCI Machine Learning Repository, 1936, doi: 10.24432/C56C76.
[45] R. Quinlan, "Thyroid Disease," UCI Machine Learning Repository, 1986, doi: 10.24432/C5D010.
[46] M. Cardoso, "Wholesale customers," UCI Machine Learning Repository, 2013, doi: 10.24432/C5030X.
[47] R. Bhatt, "Wireless Indoor Localization," UCI Machine Learning Repository, 2017, doi: 10.24432/C51880.
[48] K. Nakai, "Yeast," UCI Machine Learning Repository, 1991, doi: 10.24432/C5KG68.
[49] B. German, "Glass Identification," UCI Machine Learning Repository, 1987, doi: 10.24432/C5WW2P.

[50] R. Bock, “MAGIC Gamma Telescope,” UCI Machine Learning Repository, 2004, doi: 10.24432/C52C8B.
[51] “Heart Failure Clinical Records,” UCI Machine Learning Repository, 2020, doi: 10.24432/C5Z89R.
[52] S. Aeberhard and M. Forina, “Wine,” UCI Machine Learning Repository, 1992, doi: 10.24432/C5PC7J.
[53] A. Janosi, W. Steinbrunn, M. Pfisterer, and R. Detrano, “Heart Disease,” UCI Machine Learning Repository, 1989, doi: 10.24432/C52P4X.
[54] R. Forsyth, “Zoo,” UCI Machine Learning Repository, 1990, doi: 10.24432/C5R59V.
[55] W. Wolberg, O. Mangasarian, N. Street, and W. Street, “Breast Cancer Wisconsin (Diagnostic),” UCI Machine Learning Repository, 1993, doi: 10.24432/C5DW2B.
[56] R. A. Fisher, “The use of multiple measurements in taxonomic problems,” Ann. Eugenics, vol. 7, no. 2, pp. 179–188, Sep. 1936, doi: 10.1111/j.1469-1809.1936.tb02137.x.
[57] scikit-learn developers. “LinearDiscriminantAnalysis.” scikit-learn documentation. Accessed: May 8, 2026. [Online]. Available: https://scikit-learn.org/stable/modules/generated/sklearn.discriminant_analysis.LinearDiscriminantAnalysis.html
[58] P. Tüfekci and H. Kaya, “Combined Cycle Power Plant,” UCI Machine Learning Repository, 2014, doi: 10.24432/C5002N.
[59] T. Brooks, D. Pope, and M. Marcolini, “Airfoil Self-Noise,” UCI Machine Learning Repository, 1989, doi: 10.24432/C5VW2C.
[60] J. Gerritsma, R. Onnink, and A. Versluis, “Yacht Hydrodynamics,” UCI Machine Learning Repository, 1981, doi: 10.24432/C5XG7R.
[61] I. Yeh, “Real Estate Valuation,” UCI Machine Learning Repository, 2018, doi: 10.24432/C5J30W.
[62] A. Tsanas and A. Xifara, “Energy Efficiency,” UCI Machine Learning Repository, 2012, doi: 10.24432/C51307.
[63] I. Yeh, “Concrete Compressive Strength,” UCI Machine Learning Repository, 1998, doi: 10.24432/C5PK67.
[64] scikit-learn developers. “load_diabetes.” scikit-learn documentation. Accessed: May 8, 2026. [Online]. Available: https://scikit-learn.org/stable/modules/generated/sklearn.datasets.load_diabetes.html
[65] P. Cortez, A. Cerdeira, F. Almeida, T. Matos, and J. Reis, “Wine Quality,” UCI Machine Learning Repository, 2009, doi: 10.24432/C56S3T.
[66] P. Cortez, A. Cerdeira, F. Almeida, T. Matos, and J. Reis, “Modeling wine preferences by data mining from physicochemical properties,” Decis. Support Syst., vol. 47, no. 4, pp. 547–553, Nov. 2009, doi: 10.1016/j.dss.2009.05.016.
[67] A. Coraddu, L. Oneto, A. Ghio, S. Savio, D. Anguita, and M. Figari, “Condition Based Maintenance of Naval Propulsion Plants,” UCI Machine Learning Repository, 2014, doi: 10.24432/C5K31K.
[68] P. Händel, M. Skoglund, and M. Pettersson, “A calibration scheme for imperfect quantizers,” IEEE Trans. Instrum. Meas., vol. 49, no. 5, pp. 1063–1068, 2000, doi: 10.1109/19.872931.
[69] S. Qiu, D. Zhou, M. Kinyua, and J. Silva-Martinez, “Look-up table-based digital calibration for the correction of pipeline analog-to-digital converter non-idealities,” IEEE Access, vol. 12, pp. 196828–196841, 2024, doi: 10.1109/ACCESS.2024.3520980.
[70] IEEE Standard for Terminology and Test Methods for Analog-to-Digital Converters, IEEE Std 1241-2023 (Revision of IEEE Std 1241-2010), Oct. 2023.
[71] G. C. Valley, “Photonic analog-to-digital converters,” Opt. Express, vol. 15, no. 5, pp. 1955–1982, Mar. 2007, doi: 10.1364/OE.15.001955.
[72] Analog Devices, “Understanding noise, ENOB, and effective resolution in analog-to-digital converters,” May 7, 2012. [Online]. Available: https://www.analog.com/en/resources/technical-articles/2022/07/16/11/27/noise-enob-and-effective-resoluition-in-analog-to-digital-converter-circuits--maxim-integrated.html
[73] scikit-learn developers. “MLPClassifier.” scikit-learn documentation. Accessed: May 8, 2026. [Online]. Available: https://scikit-learn.org/stable/modules/generated/sklearn.neural_network.MLPClassifier.html
[74] Automotive Electronics Council, Component Technical Committee, “Failure mechanism based stress test qualification for integrated circuits in automotive applications,” AEC-Q100 Rev-J, Aug. 2023. [Online]. Available: http://www.aecouncil.com/Documents/AEC_Q100_Rev_J_Base_Document.pdf
[75] Automotive Electronics Council, Component Technical Committee, “Failure mechanism based stress test qualification for optoelectronic semiconductors in automotive applications,” AEC-Q102 Rev-A, Apr. 2020. [Online]. Available: https://www.aecouncil.com/Documents/AEC-Q102_Rev_A.pdf

# SUPPLEMENTARY MATERIAL

TABLE S-I

CLASSIFICATION PERFORMANCE ON BENCHMARK DATASETS COMPARING LOGREG AND VPEP+LOGREG

| **Dataset [ref.]** | **#Samples** | **#Features** | $\rho$ | **#Classes** | **Metric** | **Baseline: LogReg** | | **Proposed: VPEP + LogReg** | | **Improvement Δ (Proposed−Baseline)** |
|---|---|---|---|---|---|---|---|---|---|---|
| | | | | | | **mean** | **std** | **mean** | **std** | |
| Land Mines-1 [42],[43] | 338 | 3 | 4 | 5 | Accuracy | 0.48849 | 0.03542 | 0.65299 | 0.04453 | 0.16450 |
| | | | | | Macro-F1 | 0.45125 | 0.03516 | 0.62083 | 0.04545 | 0.16958 |
| Iris [44] | 150 | 4 | 5 | 3 | Accuracy | 0.95533 | 0.03665 | 1.00000 | 0.00000 | 0.04467 |
| | | | | | Macro-F1 | 0.95512 | 0.03686 | 1.00000 | 0.00000 | 0.04488 |
| Thyroid (New) [45] | 215 | 5 | 6 | 3 | Accuracy | 0.95395 | 0.02799 | 0.98837 | 0.01350 | 0.03442 |
| | | | | | Macro-F1 | 0.92875 | 0.04679 | 0.98253 | 0.02040 | 0.05378 |
| Wholesale Customers [46] | 440 | 7 | 3 | 3 | Accuracy | 0.71386 | 0.00880 | 0.74114 | 0.02621 | 0.02727 |
| | | | | | Macro-F1 | 0.28020 | 0.01253 | 0.39752 | 0.04962 | 0.11732 |
| Wireless Indoor Localization [47] | 2000 | 7 | 4 | 4 | Accuracy | 0.97920 | 0.00478 | 0.98720 | 0.00486 | 0.00800 |
| | | | | | Macro-F1 | 0.97921 | 0.00477 | 0.98721 | 0.00485 | 0.00801 |
| Yeast [48] | 1484 | 8 | 3 | 10 | Accuracy | 0.59212 | 0.02781 | 0.61032 | 0.02363 | 0.01820 |
| | | | | | Macro-F1 | 0.54398 | 0.05601 | 0.58220 | 0.03975 | 0.03822 |
| Glass Identification [49] | 214 | 9 | 1 | 6 | Accuracy | 0.63161 | 0.05093 | 0.73960 | 0.05559 | 0.10800 |
| | | | | | Macro-F1 | 0.54167 | 0.09130 | 0.76996 | 0.07158 | 0.22829 |
| MAGIC Gamma [50] | 19020 | 10 | 1 | 2 | Accuracy | 0.79095 | 0.00625 | 0.99097 | 0.00136 | 0.20002 |
| | | | | | Macro-F1 | 0.75679 | 0.00721 | 0.99009 | 0.00149 | 0.23329 |
| Heart Failure Clinical Records [51] | 299 | 12 | 1 | 2 | Accuracy | 0.82446 | 0.04009 | 0.83581 | 0.03475 | 0.01135 |
| | | | | | Macro-F1 | 0.79157 | 0.04883 | 0.80237 | 0.03996 | 0.01080 |
| Wine [52] | 178 | 13 | 2 | 3 | Accuracy | 0.98043 | 0.02271 | 0.98879 | 0.01598 | 0.00837 |
| | | | | | Macro-F1 | 0.98021 | 0.02265 | 0.98889 | 0.01581 | 0.00869 |
| Heart Disease (processed Cleveland) [53] | 303 | 13 | 1 | 2 | Accuracy | 0.83242 | 0.04268 | 0.84526 | 0.04233 | 0.01284 |
| | | | | | Macro-F1 | 0.82987 | 0.04334 | 0.84305 | 0.04300 | 0.01318 |
| Zoo [54] | 101 | 16 | 1 | 7 | Accuracy | 0.94754 | 0.04427 | 0.96042 | 0.03998 | 0.01288 |
| | | | | | Macro-F1 | 0.85026 | 0.12835 | 0.90639 | 0.10075 | 0.05613 |
| Breast Cancer (Diagnostic) [55] | 569 | 30 | 1 | 2 | Accuracy | 0.97698 | 0.01472 | 0.97380 | 0.01524 | -0.00318 |
| | | | | | Macro-F1 | 0.97527 | 0.01578 | 0.97182 | 0.01637 | -0.00345 |

Values are reported as means and standard deviations over repeated stratified $k$-fold cross-validation with 10 repeats using identical data splits for both methods ($k = 5$; for Zoo, $k = 4$ due to insufficient samples in a minority class for five stratified folds). Metrics are accuracy and macro-F1 (higher is better). Improvement Δ denotes the absolute gain in mean performance (Proposed − Baseline). $\rho$ denotes the feature replication factor used to map the original feature vector to the 32-input-port VPEP (effective input dimension = #Features × $\rho$); remaining unused ports were zero-padded.

TABLE S-II

REGRESSION PERFORMANCE ON 12 BENCHMARKS COMPARING LINREG AND VPEP+LINREG

| Dataset [ref.] | #Samples | #Features | $\rho$ | Target (unit) | Metric | Baseline: LinReg | | Proposed: VPEP + LinReg | | Improvement Δ (RMSE↓%, Δ$R^2$) |
|---|---|---|---|---|---|---|---|---|---|---|
| | | | | | | mean | std | mean | std | |
| Combined Cycle Power Plant [58] | 9568 | 4 | 4 | Net electrical power output (MW) | RMSE | 4.5583 | 0.0940 | 4.2293 | 0.1089 | 7.22 |
| | | | | | $R^2$ | 0.9286 | 0.0030 | 0.9385 | 0.0032 | 0.0099 |
| Airfoil Self-Noise [59] | 1503 | 5 | 4 | Scaled sound pressure level (dB) | RMSE | 4.8207 | 0.2104 | 4.1469 | 0.1711 | 13.98 |
| | | | | | $R^2$ | 0.5084 | 0.0415 | 0.6360 | 0.0330 | 0.1275 |
| Yacht Hydrodynamics [60] | 308 | 6 | 2 | Residuary resistance coefficient (unitless) | RMSE | 9.0882 | 1.1020 | 4.2223 | 0.6981 | 53.54 |
| | | | | | $R^2$ | 0.6250 | 0.0491 | 0.9186 | 0.0180 | 0.2936 |
| Real Estate Valuation [61] | 414 | 6 | 3 | House price per unit area ($10^4$ NTD/ping) | RMSE | 8.8276 | 1.4565 | 7.8562 | 1.5155 | 11.00 |
| | | | | | $R^2$ | 0.5698 | 0.0879 | 0.6577 | 0.0915 | 0.0878 |
| Energy Efficiency (Heating load) [62] | 768 | 8 | 2 | Heating load (n/a) | RMSE | 2.1047 | 0.1086 | 2.5722 | 0.1424 | -22.21 |
| | | | | | $R^2$ | 0.9560 | 0.0052 | 0.9342 | 0.0084 | -0.0218 |
| Energy Efficiency (Cooling load) [62] | 768 | 8 | 4 | Cooling load (n/a) | RMSE | 2.4313 | 0.1255 | 2.9394 | 0.1877 | -20.90 |
| | | | | | $R^2$ | 0.9341 | 0.0054 | 0.9035 | 0.0109 | -0.0306 |
| Concrete Compressive Strength [63] | 1030 | 8 | 1 | Concrete compressive strength (MPa) | RMSE | 10.4659 | 0.4423 | 8.2681 | 0.3433 | 21.00 |
| | | | | | $R^2$ | 0.6028 | 0.0363 | 0.7517 | 0.0266 | 0.1489 |
| Diabetes [64] | 442 | 10 | 1 | Disease progression (1-year, unitless) | RMSE | 54.8501 | 4.1257 | 54.0196 | 3.6499 | 1.51 |
| | | | | | $R^2$ | 0.4838 | 0.0818 | 0.4997 | 0.0724 | 0.0159 |
| Wine Quality (red) [65] | 1599 | 11 | 2 | Wine quality score (unitless) | RMSE | 0.6523 | 0.0329 | 0.6395 | 0.0310 | 1.96 |
| | | | | | $R^2$ | 0.3414 | 0.0546 | 0.3667 | 0.0530 | 0.0254 |
| Wine Quality (white) [65] | 4898 | 11 | 2 | Wine quality score (unitless) | RMSE | 0.7549 | 0.0213 | 0.7386 | 0.0186 | 2.15 |
| | | | | | $R^2$ | 0.2724 | 0.0289 | 0.3034 | 0.0228 | 0.0310 |
| Condition Based Maintenance of Naval Propulsion Plants 1 [67] | 11934 | 16 | 1 | Compressor decay coefficient kMc (unitless) | RMSE | 0.0058 | 0.0001 | 0.0036 | 0.0002 | 37.80 |
| | | | | | $R^2$ | 0.8424 | 0.0056 | 0.9389 | 0.0054 | 0.0965 |
| Condition Based Maintenance of Naval Propulsion Plants 2 [67] | 11934 | 16 | 2 | Turbine decay coefficient kMt (unitless) | RMSE | 0.0022 | 0.0000 | 0.0038 | 0.0001 | -70.57 |
| | | | | | $R^2$ | 0.9106 | 0.0028 | 0.7399 | 0.0082 | -0.1707 |

Values are reported as means and standard deviations over repeated five-fold cross-validation with 10 repeats using identical data splits for both methods. Metrics are root-mean-square error (RMSE) (lower is better) and coefficient of determination ($R^2$) (higher is better). Improvement is reported as relative RMSE reduction (%) = (Baseline − Proposed)/Baseline × 100 and absolute $R^2$ gain $\Delta R^2$ = (Proposed − Baseline). $\rho$ denotes the feature replication factor used to map the original feature vector to the 32-input-port VPEP (effective input dimension = #Features × $\rho$); remaining unused ports were zero-padded.

TABLE S-III

ENCODING CAPACITY METRICS AND ANALOG-TO-DIGITAL CONVERSION PERFORMANCE

FOR VPEP-BASED ANALOG-TO-DIGITAL CONVERSION TASK

| **L (levels) (comparator per ch)** | **Capacity metrics** | | | **Decoder** | **A/D conversion performance** | | | |
|---|---|---|---|---|---|---|---|---|
| | **K** | **H (bits)** | **$2^H$** | | **NRMSE (%FS)** | **p99 \|e\| (LSB)** | **MVR (%)** | **ENOB (equiv.)** |
| 2 (1) | 46 | 5.01 | 32.3 | LinReg | 12.13 | 286 | 6.0 | 1.25 |
| | | | | LinReg + 1D LUT (128 bins) | 7.37 | 215 | 5.6 | 1.97 |
| | | | | LinReg + 1D LUT (256 bins) | 7.16 | 215 | 5.4 | 2.01 |
| | | | | LinReg + 1D LUT (512 bins) | 6.98 | 215 | 5.4 | 2.05 |
| | | | | LinReg + 1D LUT (1024 bins) | 5.92 | 167 | 5.2 | 2.29 |
| 3 (2) | 108 | 6.19 | 73.2 | LinReg | 9.85 | 267 | 9.9 | 1.55 |
| | | | | LinReg + 1D LUT (128 bins) | 4.51 | 168 | 9.0 | 2.68 |
| | | | | LinReg + 1D LUT (256 bins) | 2.99 | 112 | 7.5 | 3.27 |
| | | | | LinReg + 1D LUT (512 bins) | 1.96 | 87 | 6.5 | 3.88 |
| | | | | LinReg + 1D LUT (1024 bins) | **0.75** | **23** | **6.1** | **5.27** |
| 4 (3) | 163 | 6.74 | 106.6 | LinReg | 7.49 | 225 | 12.2 | 1.95 |
| | | | | LinReg + 1D LUT (128 bins) | 4.11 | 142 | 11.9 | 2.81 |
| | | | | LinReg + 1D LUT (256 bins) | 3.54 | 128 | 10.9 | 3.03 |
| | | | | LinReg + 1D LUT (512 bins) | 2.45 | 122 | 9.2 | 3.56 |
| | | | | LinReg + 1D LUT (1024 bins) | 1.67 | 75 | 8.1 | 4.11 |

The table is organized from left to right to separate encoder-side capacity (decoder-independent) from decoder-side analog-to-digital conversion accuracy (decoder-dependent). $L$ denotes the number of quantization levels per photodetector channel; the number of comparators per channel is $(L - 1)$ (shown in parentheses). Capacity metrics are computed from the discrete encoder output codes after per-channel quantization and are independent of the decoder. $K$ is the number of unique codes observed over the sweep, $H$ (bits) is the Shannon entropy of the empirical code distribution, and $2^H$ is the corresponding effective number of codes. Analog-to-digital conversion metrics are reported for multiple decoders applied to the same quantized outputs. LinReg denotes a linear readout and LinReg + 1D LUT ($B$ bins) denotes a residual-correction decoder consisting of LinReg followed by a one-dimensional lookup table (LUT) with $B$ bins. Normalized root-mean-square error (NRMSE) (% of full scale, FS), p99 $|e|$ (LSB), and monotonicity violation rate (MVR) (%) are defined in the main text; ENOB denotes the quantization-noise-equivalent effective number of bits defined therein. Bold indicates the performance metrics for the configuration with the lowest NRMSE.

TABLE S-IV

CHARACTER- AND BLOCK-LEVEL RECOVERY PERFORMANCE FOR PHOTONIC INFORMATION ENCODING USING VPEP

| **Symbol set size C** | **#Data** | **Decoder** | **CER (%)** | | **BLER (%)** | |
|---|---|---|---|---|---|---|
| | | | **mean** | **std** | **mean** | **std** |
| 10 (digits) (0-9) | 10k | LDA | 6.49 | 0.25 | 22.60 | 0.89 |
| | | LogReg | 4.81 | 0.20 | 17.69 | 0.74 |
| | | MLP(1×128) | 1.96 | 0.38 | 7.19 | 1.35 |
| 26 (lowercase) (a-z) | 200k | LDA | 34.02 | 0.10 | 79.46 | 0.15 |
| | | LogReg | 29.29 | 0.12 | 73.50 | 0.18 |
| | | MLP(1×128) | 6.41 | 0.19 | 21.47 | 0.56 |
| 32 (special characters) (! # $ %...) | 200k | LDA | 51.40 | 0.12 | 94.80 | 0.11 |
| | | LogReg | 46.78 | 0.08 | 91.68 | 0.11 |
| | | MLP(1×128) | 21.44 | 0.26 | 57.43 | 0.56 |
| 52 (upper + lower) (A-Z, a-z) | 200k | LDA | 56.34 | 0.14 | 96.06 | 0.09 |
| | | LogReg | 51.17 | 0.11 | 93.01 | 0.09 |
| | | MLP(1×128) | 28.33 | 0.31 | 67.09 | 0.53 |
| 94 (printable ASCII) | 200k | LDA | 71.34 | 0.11 | 99.44 | 0.04 |
| | | LogReg | 67.19 | 0.11 | 98.65 | 0.06 |
| | | MLP(1×128) | 52.07 | 0.30 | 91.88 | 0.28 |

#Data denotes the number of evaluated four-character blocks (each block consists of four ASCII characters). Character error rate (CER) and four-character block error rate (BLER) are reported for byte-wise decoding under different symbol set sizes $C$, ranging from digits ($C = 10$) to printable ASCII excluding space ($C = 94$). Results are shown for three decoders: LDA, LogReg, and MLP (1 × 128). Values are reported as means and standard deviations over repeated five-fold cross-validation with five repeats using identical data splits across decoders. CER is evaluated at the character level and BLER denotes the fraction of four-character blocks containing at least one incorrectly recovered character (lower is better).